%% file: Main.tex
\documentclass[sigconf]{acmart}
\usepackage{multirow}
\usepackage{array}
\usepackage{graphicx} % 이미지 포함을 위한 패키지
\usepackage{float} % 수동 위치 지정을 위한 패키지
\usepackage{algorithm}
\usepackage{algorithmic}
\usepackage{xspace}
\usepackage{graphicx}
\usepackage{etoolbox}
\usepackage{float}
\usepackage[utf8]{inputenc}
\usepackage{xcolor}
\usepackage{balance}  % for  \balance command ON LAST PAGE  (only there!)
\usepackage{hyperref}
\usepackage{booktabs}
\usepackage{tabularx}
\usepackage{caption}
\usepackage{mathtools}
\usepackage{url}
\usepackage{amsfonts}
\usepackage{algorithmic}
\usepackage{textcomp}
\usepackage{algorithm}
\usepackage{multirow}
\usepackage{color}
\usepackage{soul}
\usepackage{subcaption}
\usepackage{enumitem}
\usepackage{array}

\newcolumntype{L}[1]{>{\raggedright\let\newline\\\arraybackslash\hspace{0pt}}m{#1}}
\newcolumntype{C}[1]{>{\centering\let\newline\\\arraybackslash\hspace{0pt}}m{#1}}
\newcolumntype{R}[1]{>{\raggedleft\let\newline\\\arraybackslash\hspace{0pt}}m{#1}}

\usepackage[normalem]{ulem}
\usepackage{color}
\usepackage{xcolor}

\AtBeginDocument{%
  }

\setcopyright{acmlicensed}
\copyrightyear{2025}
\acmYear{2025}
\acmDOI{XXXXXXX.XXXXXXX}

\acmConference[Conference acronym 'XX]{Make sure to enter the correct
  conference title from your rights confirmation emai}{June 03--05,
  2025}{Woodstock, NY}
\acmISBN{978-1-4503-XXXX-X/18/06}

\newcommand{\system}{\emph{HAPPIER}\xspace}

\newcommand{\foneone}{\emph{Interaction graph}\xspace}
\newcommand{\fonetwo}{\emph{Impact search}\xspace}
\newcommand{\fonethree}{\emph{Docking simulation}\xspace}
\newcommand{\ftwo}{\emph{PPI-Bookmark}\xspace}

\newcommand{\panone}{\emph{PPI-Graph Panel}\xspace}
\newcommand{\pantwo}{\emph{Detail Panel}\xspace}

\makeatletter

\newcommand{\RomanNum}[1]{\uppercase\expandafter{\romannumeral #1}}

\begin{document}

%%
%% The "title" command has an optional parameter,
%% allowing the author to define a "short title" to be used in page headers.
% \title{Enhancing Hypothesis Generation for Target Identification in Drug Discovery with \system: Facilitating Iterative Cycles of Divergent and Convergent Thinking}

\title[Supporting Medicinal Chemists in Iterative Hypothesis Generation for Drug Target Identification]{Supporting Medicinal Chemists in Iterative Hypothesis Generation for Drug Target Identification}

\author{Youngseung Jeon}
\email{ysj@ucla.edu}
\affiliation{%
  \institution{University of California}
  \city{Los Angeles, CA}
  \country{USA}
}

\author{Christopher Hwang}
\email{chrishwang42@ucla.edu}
\affiliation{%
  \institution{University of California}
  \city{Los Angeles, CA}
  \country{USA}
}

\author{Ziwen Li}
\email{zil105@g.ucla.edu}
\affiliation{%
  \institution{University of California}
  \city{Los Angeles, CA}
  \country{USA}
}

\author{Taylor Le Lievre}
\email{tlelievr@purdue.edu}
\affiliation{%
  \institution{Purdue University}
  \city{West Lafayette, IN}
  \country{USA}
}

\author{Jesus J. Campagna}
\email{JCampagna@mednet.ucla.edu}
\affiliation{%
  \institution{University of California}
  \city{Los Angeles, CA}
  \country{USA}
}

\author{Cohn Whitaker}
\email{wcohn@usc.edu}
\affiliation{%
  \institution{University of Southern California}
  \city{Los Angeles, CA}
  \country{USA}
}

\author{Varghese John}
\email{VJohn@mednet.ucla.edu}
\affiliation{%
  \institution{University of California}
  \city{Los Angeles, CA}
  \country{USA}
}

\author{Eunice Jun}
\email{emjun@ucla.edu}
\affiliation{%
  \institution{University of California}
  \city{Los Angeles, CA}
  \country{USA}
}

\author{Xiang `Anthony' Chen}
\authornote{Corresponding author}
\email{xac@ucla.edu}
\affiliation{%
  \institution{University of California}
  \city{Los Angeles, CA}
  \country{USA}
}

%%
%% By default, the full list of authors will be used in the page
%% headers. Often, this list is too long, and will overlap
%% other information printed in the page headers. This command allows
%% the author to define a more concise list
%% of authors' names for this purpose.
\renewcommand{\shortauthors}{Jeon et al.}

\begin{abstract} %150 words
While drug discovery is vital for human health, the process remains inefficient. Medicinal chemists must navigate a vast protein space to identify target proteins that meet three criteria: physical and functional interactions, therapeutic impact, and docking potential. Prior approaches have provided fragmented support for each criterion, limiting the generation of promising hypotheses for wet-lab experiments. We present \system, an AI-powered tool that supports hypothesis generation with integrated multi-criteria support for target identification. \system enables medicinal chemists to 1) efficiently explore and verify proteins in a single integrated graph component showing multi-criteria satisfaction and 2) validate AI suggestions with domain knowledge. These capabilities facilitate iterative cycles of divergent and convergent thinking, essential for hypothesis generation. We evaluated \system with ten medicinal chemists, finding that it increased the number of high-confidence hypotheses and support for the iterative cycle, and further demonstrated the relationship between engaging in such cycles and confidence in outputs.

\end{abstract}

\begin{teaserfigure}
        \centering
    \includegraphics[width=0.90\textwidth]{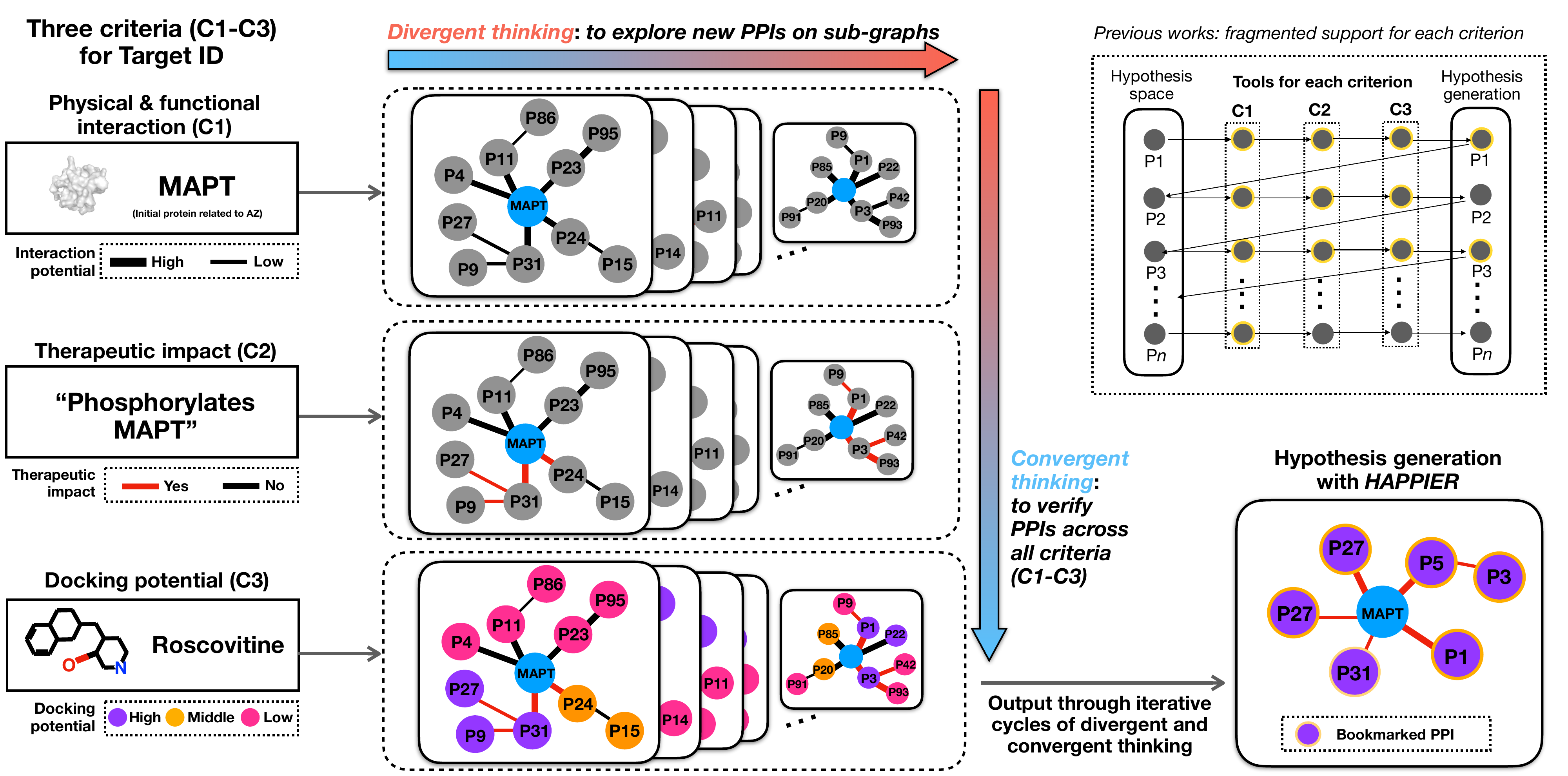}
    \caption{\system helps medical chemists efficiently generate hypotheses with integrated multi-criteria support in identifying drug targets (proteins) to satisfy all criteria (C1-C3). Beyond prior systems that offer fragmented support for each criterion, \system allows medicinal chemists to simultaneously explore and verify hundreds of hypothesis candidates across all criteria on the integrated graph-based visualization. 
    This facilitates iterative cycles of divergent and convergent thinking, which are essential for hypothesis generation. 
    }
    \label{fig:head}
\end{teaserfigure}

\maketitle
\makeatother



%%
%% Keywords. The author(s) should pick words that accurately describe
%% the work being presented. Separate the keywords with commas.


% \received{20 February 2007}
% \received[revised]{12 March 2009}
% \received[accepted]{5 June 2009}

%%
%% This command processes the author and affiliation and title
%% information and builds the first part of the formatted document.

\input{text/01_Intro.tex}

\input{text/02_Walkthrough.tex}

\input{text/03-1_Rel}
\input{text/04_Formative_study}

\input{text/05_HAPPIER}

\input{text/06_User_Study}

\input{text/08_Results}

\input{text/09_Discussion}

\input{text/10_Conclusion}

%%
%% The next two lines define the bibliography style to be used, and
%% the bibliography file.

\bibliographystyle{ACM-Reference-Format}
\bibliography{Reference}

\clearpage
\appendix
\input{text/Appendix}
\clearpage
\end{document}

%% file: text/01_Intro.tex
\section{Introduction}
Drug discovery is one of the most important domains to improve human health, addressing diseases that threaten human life. For example, Zidovudine (AZT), the first antiretroviral drug for HIV/AIDS, marked a turning point in the fight against the global epidemic and laid the foundation for saving lives~\cite{broder2010development}. However, the discovery of a new drug involves multiple phases of screening, synthesizing, and testing, a process that can span over 14 years and cost over \$1 billion~\cite{wouters2020estimated, hinkson2020accelerating}.

One of the key early steps in this time- and cost-intensive process is identifying target proteins for drugs to bind to, a process called \textit{target identification (Target ID)}. Target ID is the process of finding a target protein with which a drug can be docked. 
When a drug binds to a target protein (TP), TP affects an initial protein (IP) that is therapeutically related to a specific disease, thereby treating the disease (Please see definitions of terminologies in {\S}\ref{back}). Medicinal chemists must generate a list of target proteins (hypotheses)—within the vast protein target space—estimated to range from 20,000~\cite{adkins2002toward} to several hundred thousand~\cite{smith2013proteoform}—for wet-lab validation that meet three key criteria: C1) physical and functional interactions with the IP, C2) therapeutic impact on the IP, and C3) docking potential with a reference drug.
Medicinal chemists must be highly selective in generating hypotheses about a target protein, because wet-lab validation is expensive (around \$1400 per protein).
% ~\footnote{Due to anonymity concerns, the reference cannot be disclosed at this stage.})
As a result, hypothesis generation across such a vast protein space makes Target ID time-consuming and costly, hindering progress in human health.

To expedite hypothesis generation in drug discovery, scientific discovery support tools have been proposed to validate each individual criterion in Target ID (Table~\ref{tab:SDtool} in the appendix), e.g., \href{https://string-db.org/}{STRING} for C1, \href{https://scholar.google.com/}{Google Scholar} for C2, and \href{http://www.swisstargetprediction.ch}{SwissTarget Prediction} for C3. While these tools each support an individual criterion during Target ID, currently no single tool integrates {\it all} three criteria in one cohesive platform. 
This fragmented support prevents medicinal chemists from engaging in the cycles of divergent and convergent thinking essential for hypothesis generation~\cite {klahr1988dual,schickore2014scientific,schunn19954}: divergent thinking involves broadly exploring ideas to expand the hypothesis space beyond existing knowledge; convergent thinking narrows this space by evaluating candidates against multiple criteria; and iterative cycles of divergence and convergence progressively expand and refine the space to identify a few reliable candidate hypotheses.

Such a gap of effective tool support for Target ID presents a unique opportunity for HCI where a design centered on domain experts can potentially lead to an integrated interface that enables medicinal chemists to efficiently generate hypotheses across multiple scientific criteria. 
This, in turn, can result in accelerating scientific discovery in drugs and offering design principles for AI-powered tools in other health domains such as bioinformatics~\cite{karim2023explainable, lin2025bridging}, clinical decision support~\cite{weiner2022effect, elhaddad2024ai}, and public health~\cite{jungwirth2023artificial, olawade2023using}.

To understand Target ID, we conducted a formative study with five medicinal chemists and identified three major challenges in the Target ID workflow: the complexity of exploring large protein-protein interactions (PPIs) graphs when hundreds of PPIs are presented at once, which hinders exploring new potential candidates (divergent thinking), the inefficiency of evaluating these criteria using their existing tools, which prevents narrowing down candidates (convergent thinking), and the absence of an integrated tool that enables the exploration and verification of promising hypotheses (iterative cycles of divergent and convergent thinking).

To tackle the challenges in the Target ID workflow, we present \system (Human-AI Protein-Protein Interaction discovERy), an AI-powered interface for efficient hypothesis generation in Target ID. \system enables medicinal chemists to seamlessly both explore (i.e., divergent thinking) and verify (i.e., convergent thinking) proteins across all three criteria (C1–C3) through an integrated graph-based visualization (Figure~\ref{fig:head}).
We build a unified interface that facilitates the iterative cycle of divergent and convergent thinking in Target ID by integrating AI models into medicinal chemists’ workflows.
Furthermore, we investigate the impact of engaging experts in an iterative cycle of divergent and convergent thinking on their perceived confidence for hypothesis outputs to understand how \system facilitates this process.
Specifically, we investigate these two research questions (RQs):

\begin{itemize} [leftmargin=0.25in]
    \item RQ1: How can we design and develop an AI-powered integrated interface that supports medicinal chemists in the hypothesis generation process for the Target ID task?
    \item RQ2: Does expert engagement in both divergent and convergent thinking significantly affect their perceived confidence in hypothesis output?
\end{itemize}

To evaluate \system, we conducted a user study with ten medicinal chemists. For RQ1, our findings showed significant differences in PPI outputs (i.e., the number of submitted outputs and the level of perceived confidence) and support the iterative cycles of divergent and convergent thinking between the experimental (\system) and control (existing interfaces) groups ($p < .05$). For RQ2, we identified the divergent and convergent thinking processes that occurred in \system during the task based on Linkography~\cite{goldschmidt2014linkography}, which is a method that analyzes connections between ideas and enables the observation of divergent and convergent thinking.
We categorized participants' final submitted outputs into three groups: 1) Both-DC: a protein-protein interaction was visited in both divergent and convergent processes, 2) Either-DC: in either one of the two processes, and 3) Neither-DC: in neither of them. Our findings show significant differences in perceived relevance and confidence for submitted outputs between three groups. \textit{Both-DC} showed a higher confidence level in the submitted PPIs than both \textit{Either-DC} ($p < .001$) and \textit{Neither-DC} groups ($p < .001$). This finding demonstrates how \system fosters iterative engagement in divergent and convergent thinking processes, which in turn may lead to the generation of high-confidence hypotheses.

This research's contributions include
\begin{itemize} [leftmargin=0.25in]
\vspace{-0.1cm}
    \item a tool contribution of an integrated system, \system, which integrates AI to provide tool support for medicinal chemists' workflows and enhances their ability to generate hypotheses in Target ID;
    \item experimental evidence that shows how engaging in divergent and convergent thinking facilitates high-confidence hypothesis generation; and
    \item a discussion of design implications for integrating AI and HCI for scientific discovery beyond drug discovery across health domains.
\end{itemize}

%% file: text/02_Walkthrough.tex
\section{Usage Scenario}
Below, we walk through how a scientist, Jo, uses \system in her research on drugs for Alzheimer's disease. This usage scenario is motivated by a real-world use case from our collaborators. To start, we first provide explanations of key terminology used in the scenario.

\subsection{Background \& Definitions of Key Terminology}\label{back}

\textbf{Target ID.} The goal of Target ID is to find 
target protein (TP) that can bind with the ligands they have.
When a drug binds to a target protein (TP), it interacts with an initial protein (IP) associated with the disease, thereby producing the desired therapeutic effect and treating the disease.
The reason for not selecting IP as TP and instead finding another protein is that inhibition or promotion through PPIs from another protein to IP allows for more precise control of regulation compared to direct targeting of the initial protein ~\cite{fry2015targeting, holohan2013cancer}.
\textbf{Hypothesis Generation.} In the process, the type of candidate hypothesis is protein–protein interactions (PPIs) linking the initial protein (IP) to a potential target protein (TP). Medicinal chemists generate multiple promising candidates for wet-lab validation.  
Since wet-lab validation is expensive (\$1,400 per protein), it is crucial to generate hypothesis candidates that have a high likelihood of success.

\textbf{Key Terminology.} Five terminologies are essential to Target ID: an initial protein (IP), a target protein (TP), and three criteria (C1-C3) in Target ID (Table~\ref{tab:terms}).

\begin{center}
    \input{table/Background}
\end{center}

\begin{figure*}
    \centering
    \includegraphics[width=0.9\linewidth]{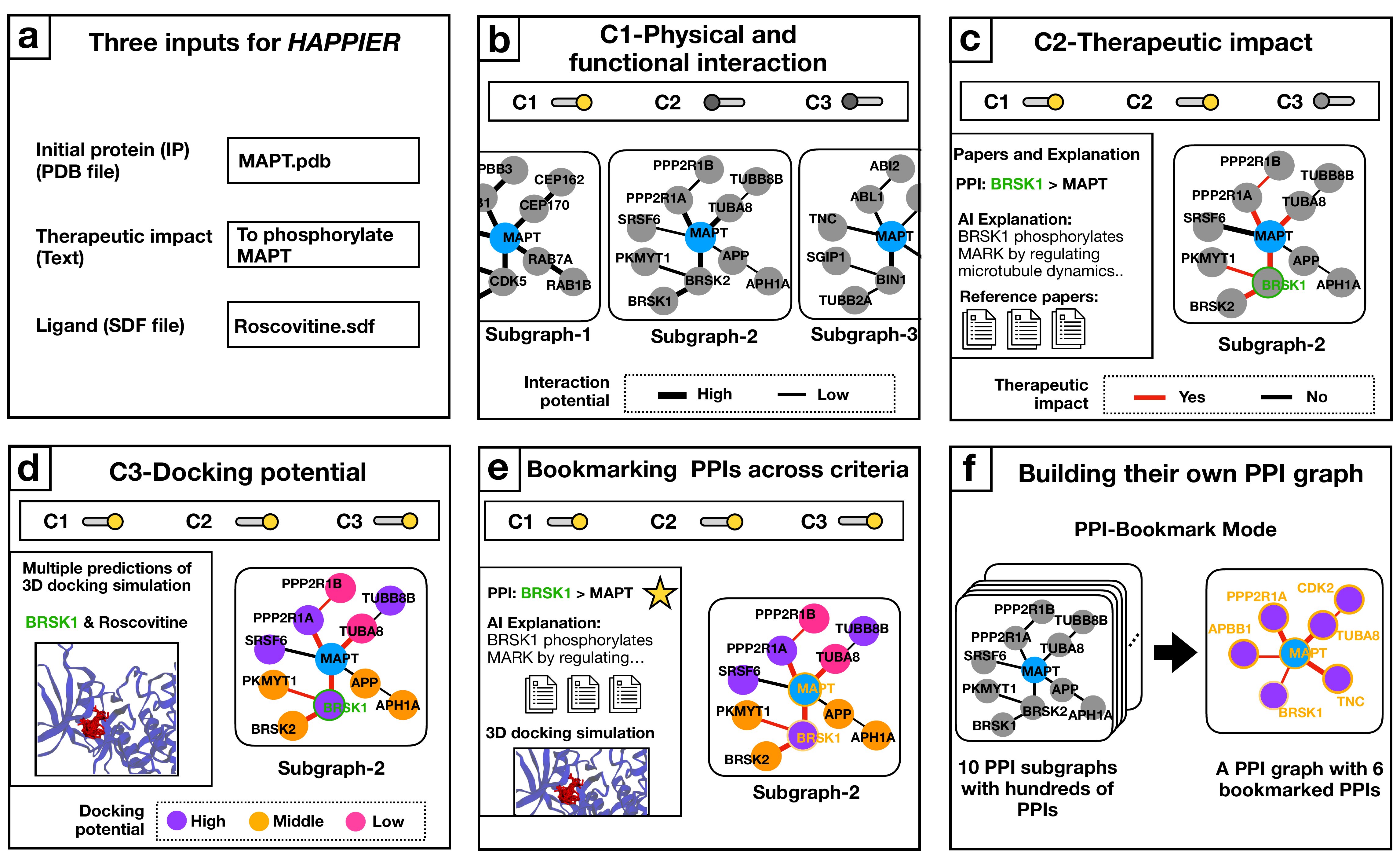}
    \caption{Usage scenario of \system to support scientific discovery in Target ID. Users input an initial protein, a therapeutic impact, and a ligand (a). They then verify protein-protein interactions (PPIs) on graphs across all three criteria (C1, C2, and C3) through AI models (b-d). Users bookmarked PPIs that are likely to satisfy all criteria (e). Users generate their own PPI graph by repeating the previous steps (f).
    }
    \label{fig:scenario}
\end{figure*}

\subsection{A Walkthrough of Interacting with \system}

Jo, a drug discovery researcher, is identifying target proteins (TP) related to MAPT, a key initial protein (IP) involved in Alzheimer’s disease. 
Because wet-lab experiments are expensive, Jo should generate novel hypotheses with a high likelihood of success that satisfy the three criteria for wet-lab validation. 
Here, the type of candidate hypothesis is protein–protein interactions (PPIs) linking the initial protein (IP, e.g., MAPT) to a potential target protein (TP, e.g., other proteins).

To begin, Jo opens \system and inputs three types of information related to the three criteria (Figure~\ref{fig:scenario}-a): (1) the initial protein, using a PDB file of MAPT\footnote{https://www.rcsb.org/structure/8P34}; (2) the therapeutic impact, specified as ``to phosphorylate MAPT''; and (3) the initial ligand, using an SDF file of Roscovitine\footnote{https://www.selleckchem.com/products/Roscovitine.html}.

First, she considers physical and functional interactions (Figure~\ref{fig:scenario}-b). She sees a subgraph centered on MAPT and then explores ten subgraphs 
By moving the slider for C1 at the top, she quickly observes that edge thickness reflects the interaction potential between two proteins (PPI), which is calculated based on how many related works support the PPI. The order of subgraphs is determined by this interaction potential.
Jo intuitively explores the full spectrum of interactions, from well-studied to less-characterized but potentially significant PPIs.

Second, Jo looks into the therapeutic impact by activating the therapeutic impact search (Figrue~\ref{fig:scenario}-c) by moving the slider for C2 at the top. She easily identifies PPIs related to 'phosphorylation of MAPT' on the graphs, which are highlighted in red. 
Since AI generates these results, she wants to validate them. She clicks on one of the PPIs, MAPT → BRSK1, to review the AI-generated explanation and supporting references. After verifying that the explanation is consistent with the cited papers, which state that BRSK1 phosphorylates MAPT by regulating microtubule dynamics, she gains confidence in the AI's output. 

Third, Jo explores the docking potential of proteins in graphs by moving the slider for C3 at the top (Figure~\ref{fig:scenario}-d). Purple nodes indicate proteins with a high binding affinity for the ligand, determined by a spatial-based docking simulation AI model. Jo confirms that BRSK1 also demonstrates strong docking potential with the Roscovitine she input. Jo recognizes that the PPI has potential because the multiple docking simulation results of the protein consistently show similar poses across all appropriate binding sites.

Satisfied with the evidence across all three criteria—physical/functional interaction, therapeutic impact, and docking potential—Jo bookmarks the PPI BRSK1 > MAPT as a strong candidate (hypothesis) for further investigation (Figure~\ref{fig:scenario}-e). Jo also explores other promising PPIs suggested by the system, validating them through the same multi-step process. Finally, Jo builds a personalized PPI graph centered around MAPT and her bookmarked PPIs, allowing her to visualize and consolidate the multiple hypotheses she has generated from hundreds of PPI candidates (Figure~\ref{fig:scenario}-f).

%% file: table/Background.tex
\begin{table*}[ht]
\centering
\caption{Key Terminology for Target Identification (Target ID).}
\renewcommand{\arraystretch}{1.3}
\begin{tabular}{p{3cm} p{11cm}}
\toprule
\textbf{Terminology} & \textbf{Description} \\
\midrule
Initial protein (IP) & The protein that therapeutically affects the disease. For example, MAPT is a crucial protein in Alzheimer’s disease treatment as its abnormal hyperphosphorylation disrupts the stability of neuron structures and leads to the buildup of tangled proteins in brain cells~\cite{strang2019mapt}. \\
Target protein (TP) & The protein that is considered a promising candidate for therapeutic intervention on a disease and should satisfy the three criteria (C1–C3) below. \\
\midrule
C1 – Physical and functional interactions & TP should have physical and functional interactions with IP, meaning that a biological action occurs, such as binding, which directly influences the function or state of the IP~\cite{stelzl2005human}. For example, BRSK1 interacts with MAPT both through direct binding and by influencing its biological function. \\
C2 – Therapeutic impact & Modulation of TP by a drug (ligand) should be involved in either activating or inhibiting a therapeutic effect related to the disease associated with IP. For example, the phosphorylation of MAPT is a major contributing factor to Alzheimer’s disease~\cite{strang2019mapt}. Inhibiting the phosphorylation of MAPT can serve as a potential therapeutic intervention. BRSK1 induces phosphorylation of MAPT; therefore, inhibiting BRSK1 may prevent MAPT hyperphosphorylation. \\
C3 – Docking potential & TP should have a high docking potential with the reference ligand. For example, BRSK1 has high docking potential with Roscovitine, one of the representative drugs for Alzheimer’s disease. Therefore, inhibiting BRSK1 through docking with Roscovitine may reduce the hyperphosphorylation of MAPT. 
% Typically, the reference ligand is later augmented through the ligand optimization process~\cite{sun2021discovery}.
\\
\bottomrule
\end{tabular}
\label{tab:terms}
\end{table*}

%% file: text/03-1_Rel.tex
\section{Related work}
Our work, as a hypothesis generation supporting tool for Target ID that facilitates both divergent and convergent thinking, builds on three major lines of previous research: (1) divergent and convergent thinking in hypothesis generation, (2) hypothesis generation supporting tools in HCI, and (3) tools and AI for Target ID.

\subsection{Divergent \& Convergent Thinking in Hypothesis Generation}

Hypothesis generation involves the process of uncovering new knowledge through systematic observation for conducting an experiment~\cite{schickore2014scientific}. 
Klahr and Dunbar~\cite{klahr1988dual} proposed the dual-space model of scientific discovery in which scientific activity occurs in two spaces: 1) a hypothesis space: hypotheses about causal relations are drawn in the current representation, and 2) an experiment space: researchers validate a hypothesis within a given problem definition. Schunn and Klahr~\cite{schunn19954} expand the two-space model to a four-space model by subdividing the hypothesis space into representation and hypothesis spaces, and the experimental space into paradigm and experiment spaces. 
The core objective in these models is to expand the hypothesis space with high confidence before transitioning into the experiment space~\cite{klahr1988dual, schunn19954, schickore2014scientific}. This suggests that generating as many plausible hypotheses as possible increases the probabilistic chance of making novel scientific discoveries.

To formulate high-confidence hypotheses, cognitive psychology research suggests that effective hypothesis generation involves several steps, each of which can require significant mental effort~\cite {klahr1988dual,schickore2014scientific,schunn19954}. First, \textbf{divergent thinking} involves broadly exploring ideas to expand the hypothesis space beyond existing knowledge. Schunn and Klahr~\cite{schunn19954} mentioned that conducting an exhaustive search of available datasets is beneficial for extracting insightful features by thoroughly exploring scientific facts. 
Second, \textbf{convergent thinking} narrows down this space by evaluating candidates against multiple criteria. Dunbar~\cite{dunbar1993concept} mentioned that efficient focus on specific types or characteristics of data through verification can be an essential step in scientific discovery.
Third, hypothesis generation involves \textbf{iterative cycles of divergence and convergence}, progressively refining and filtering the expanded space to identify a few reliable candidate hypotheses. Supporting all divergent thinking, convergent thinking, and their iterative cycle facilitates the generation of hypotheses with a high level of confidence~\cite {klahr1988dual,schickore2014scientific,schunn19954,okada1997collaborative}. 
Sawyer~\cite{sawyer2013zig} emphasizes the importance of engaging in both processes of divergent and convergent thinking, postulating that the development of a creative idea is not a linear process through either convergent or divergent thinking; rather, it follows a non-linear process in which the two modes switch repeatedly. 

In this view, ideal supporting hypothesis generation means enabling iterative cycles in which multiple hypothesis candidates are continuously explored and validated against domain-specific criteria, rather than offering fragmented assistance.
Therefore, facilitating the iterative cycles of divergent and convergent thinking is key in designing tools for hypothesis generation.

\subsection {Hypothesis Generation Support Tools in HCI}
Human-computer interaction (HCI) has developed tools for hypothesis generation, even though no studies directly support Target ID. To facilitate hypothesis generation, research focuses on supporting: 1) data retrieval for divergent thinking and 2) dataset analysis for convergent thinking.

In the case of data retrieval, previous work has introduced AI-based interfaces designed to support data review by helping users retrieve papers and datasets efficiently, which allows them to explore their hypothesis space efficiently.
They recommended papers by analyzing the researchers' data, including the papers and comments collected during their research, to support the discovery of information that had not been previously considered. 
Portenoy et al. \cite{portenoy2022bursting} suggested Bridger, which is an AI-based interface to provide a faceted representation of authors with information gleaned from their papers and inferred author personas.
PaperWeaver~\cite{lee2024paperweaver} is an enhanced paper alert system that provides contextual summaries of recommended papers based on user-collected papers using large language models (LLMs). 
DiscipLink~\cite{zheng2024disciplink} is an interactive system that uses LLMs to support interdisciplinary information-seeking by generating exploratory questions tailored to researchers' interests and facilitating paper retrieval and synthesis. 
Kang et al. \cite{kang2023comlittee} introduced ComLittee, a literature discovery system that supports author-centric exploration where users can capture and utilize users’ evolving knowledge of authors. 

In the case of data analysis, previous work has introduced AI-based interfaces designed to help researchers explore and analyze experimental datasets, which enables them to narrow down their hypothesis space efficiently.
They recommended papers by analyzing the researchers' data, including the papers and comments collected during their research, to support the discovery of information that had not been previously considered. 
These studies enable users to have intuitive and interactive access to complex scientific domain knowledge, ultimately helping refine their ideas through scientific validation. FathomGPT~\cite{khanal2024fathomgpt} is an AI-based interface that enables researchers to explore and analyze ocean science datasets where they search for images, taxonomy, and habitats of ocean species via large language models. 
MedChemLens~\cite{shi2022medchemlens} extracts molecular features from chemical literature and clinical trials, organizes them by drug structure, and visualizes key experimental factors interactively.

However, prior work primarily supports either divergent or convergent thinking in hypothesis generation, which is a fragmented support for hypothesis generation. 
We believe tackling the challenge of hypothesis generation presents a unique opportunity for HCI because it requires an integrated interface that enables medicinal chemists to generate hypotheses across multiple scientific criteria efficiently, accelerating scientific discovery in drug discovery and offering design principles for AI-powered tools in other health domains such as bioinformatics~\cite{karim2023explainable, lin2025bridging}, clinical decision support~\cite{weiner2022effect, elhaddad2024ai}, and public health~\cite{jungwirth2023artificial, olawade2023using}.
To this end, we present \system, an AI-powered tool to support hypothesis generation designed to facilitate an iterative cycle of divergent and convergent thinking to help medicinal chemists in Target ID.

\subsection {AI technologies and Platforms for Target ID} 
In the drug development field, there are existing web-based tools
for the three criteria (C1-C3) for Target ID that scientists have used (Table~\ref{tab:SDtool} in the appendix). These interfaces have separately supported these three criteria for target ID. For example, for C1, STRING~\cite{szklarczyk2021string} provides researchers with PPIs, which include physical contact between proteins through experimental results. 
For C2, Google Scholar is a widely used academic search engine that provides access to literature, including journal articles, theses, conference papers, books, and patents. It enables researchers to search by protein and desired therapeutic impact 
For C3, SwissTarget Prediction~\cite{gfeller2014swisstargetprediction} suggests lists of proteins based on molecular fingerprints of the initial proteins. This tool identifies proteins whose docking ligands are structurally similar to a given ligand.
 
Meanwhile, there is existing AI support for Target ID. 
For example, Nvidia BioNeMo~\cite{bionemo2024} is an AI platform for drug discovery that simplifies and accelerates the building and training of models using your own data and scaling the deployment of models for drug discovery applications by providing multiple AI models, such as AlphaFold~\cite{jumper2021highly}, DiffDock~\cite{corso2022diffdock}, MegaMolBART~\cite{irwin2022chemformer}, and so on. Schrödinger's Maestro~\cite{MaestroS2024} employs machine learning to predict how a molecule will fold, the resultant 3D shape, and how it might interact with other molecules. 
However, these existing tools and models often lack comprehensive integration that would allow seamless exploration and validation of PPIs with all three criteria within a single interface, which is the main design consideration as we developed \system. 

%% file: text/04_Formative_study.tex
\section{Formative study}\label{fs}
We interviewed five professional medicinal chemists to understand the existing process in Target ID, challenges, and possible solutions for these difficulties. Based on these insights, we derived design goals for \system. 

\subsection{Approach}
Over a two-month period (February–April 2024), we met with five medicinal chemists from four institutions, twice each (ten meetings in total), to understand the Target ID process.
Table~\ref{tab:demo_f} in the appendix shows demographic information of the participants. 
Their work experience ranges from 7 to 18 years (mean=11.4, SD=4.3). Each interview took approximately 60-90 minutes. Three authors attended all the meetings, and each took notes on the interviews. 
In the first interview, we asked about (1) medicinal chemists’ current workflows and criteria for Target ID and (2) the barriers and challenges that impede scientific discovery. In the second interview, we asked about (3) potential solutions and coping strategies to address these challenges. After completing the interviews, we applied thematic analysis and iterative open coding~\cite{clarke2015thematic} to analyze the interview transcripts. Three authors coded and analyzed the transcripts for emerging themes, and the findings were discussed among the co-authors iteratively until consensus was reached.

\subsection{Findings}
Below, we summarize the Target ID process and criteria, current challenges, and possible solutions that emerged from the interviews. 

\subsubsection{Target ID process and criteria}
Our findings identified the Target ID process of searching for possible protein-protein interactions (PPIs) consisting of a target protein and an initial protein. Medicinal chemists identify PPI candidates by first retrieving proteins that directly interact with the initial protein (C1: physical and functional interactions via STRING), then narrowing candidates based on therapeutic relevance (C2: filtering for disease-linked proteins) and docking potential (C3: assessing ligand–protein binding feasibility).

Medicinal chemists identify PPI candidates based on three criteria (C1, C2, and C3). They extend the scope of PPIs with C1 while constraining the scope using C2 and C3.
First, to identify physical and functional interactions (C1), scientists retrieve a PPI graph consisting of hundreds of proteins based on the initial protein on STRING.
STRING ranks PPIs by their combined score, which indicates the likelihood of PPIs existing. Rather than focusing solely on high scores, researchers aim to explore a wide range of underexplored PPIs. At this stage, they typically examine at least 200 proteins to maximize the chance of identifying a promising target. 
Second, to look for possible therapeutic impacts (C2), medicinal chemists investigate the therapeutic impacts of proteins on the graph through Google search. For example, MAPT is linked to Alzheimer’s due to its excessive phosphorylation. Scientists filter the graph to keep only proteins that phosphorylate MAPT. Third, to check the docking potential (C3), scientists identify the docking potential of proteins with their ligands by entering their ligand into a web-based interface, such as SwissTargetPrediction~\footnote{http://www.swisstargetprediction.ch/}. The tool recommends proteins whose ligands have similar structures to the ligand, providing a similarity score. Consequently, they use C2 and C3 to narrow down the list of PPIs on the graph (C1).
Researchers then compile and summarize their findings manually using MS Office\footnote{https://www.microsoft.com/en-us/microsoft-365/microsoft-office}. Medicinal chemists mentioned that this process typically takes about a month to generate a final shortlist of 5 to 20 PPIs from hundreds of initial candidates for wet-lab validation.

\subsubsection{Challenges and possible solutions}

We also identified three challenges, as well as possible solutions to address these challenges. 
First, for physical and functional interactions (C1), medicinal chemists face challenges due to the complexities in exploring novel PPIs that are less studied. 
The existing interface first presents well-studied PPIs with high combined scores, while less-studied proteins with medium or low scores are shown later in an accumulated manner.
Medicinal chemists should explore dozens to hundreds of less-studied proteins in a single screen simultaneously, hindering the exploration of new proteins. As a possible solution, they suggested creating multiple subgraphs centered on an initial protein, each ideally containing 50 to 60 proteins, so that they can focus on a manageable set of candidates while still preserving diversity and the chance of discovering underexplored PPIs.

Second, medicinal chemists face difficulties due to the repetitive process of therapeutic impact search (C2) and docking simulation (C3) across multiple proteins. They should validate a large number of proteins on the graph from STRING one by one.
They individually search for proteins on Google with their desired therapeutic impact, while simultaneously checking docking potential between proteins and a reference ligand one by one in SwissTargetPrediction.
This time-consuming process prevents medicinal chemists from identifying the most promising targets. 
As a possible solution, medicinal chemists sought to investigate the therapeutic impact and perform docking simulations on multiple protein candidates simultaneously. Additionally, they highlighted the potential of AI to support this process and emphasized the necessity of domain knowledge to justify AI suggestions, such as related papers for therapeutic impact search (C2) and multiple docking cases for docking simulation (C3). They noted that AI’s performance could be considered adequate when multiple docking cases are positioned in locations deemed appropriate.

Lastly, medicinal chemists struggle due to the absence of integrated tools where they can explore and verify PPI candidates across all criteria (C1-3). 
Manually organizing the PPIs validated through multiple interfaces for the three criteria in MS Word and Excel is time-consuming, tedious, and prone to errors. This limitation prevents them from having a sufficient number of PPIs included, which leads to medicinal chemists struggling to refine their ideas through iterative cycles of exploring and verifying PPIs. 
As a possible solution, medicinal chemists strongly desired an integrated interface to organize PPIs with all criteria satisfaction in a single view.

In summary, medicinal chemists highlighted three key challenges that hinder scientific discovery: the difficulty of exploring less-studied PPIs (C1), the inefficiency of repetitive therapeutic impact searches and docking simulations (C2–C3), and the absence of an integrated interface to organize and compare results across all criteria (C1-3). To address these challenges, they envisioned an integrated tool that (1) supports subgraph-level exploration, (2) enables simultaneous therapeutic impact search and docking simulation across multiple proteins, and (3) provides a unified platform to show multi-criteria satisfaction of PPI candidates.

\subsection{System Design Goals}
Based on the findings, we established three key design goals for an AI-powered interface to support Target ID. 
\begin{itemize}
\item\textbf{Goal 1 - Divergent thinking}: to provide multiple subgraphs, \textit{i.e.,} different versions of physical and functional interaction graphs based on interaction likelihood to balance manageability and diversity. This expands the scope of potential PPI candidates efficiently.
\item\textbf{Goal 2 - Convergent thinking}: to provide AI suggestions for therapeutic impact-based search and docking simulations with multiple PPI inputs, while simultaneously offering supporting information to verify these suggestions. This ultimately narrows the scope.
\item\textbf{Goal 3 - Iterative cycles of divergent and convergent thinking}: to enable exploration and verification of PPIs that satisfy multiple criteria within a single integrated view. This promotes iterative refinement of the scope by expanding and narrowing it. 
\end{itemize}

%% file: text/05_HAPPIER.tex
\begin{figure*}
    \centering
    \includegraphics[width=1\linewidth]{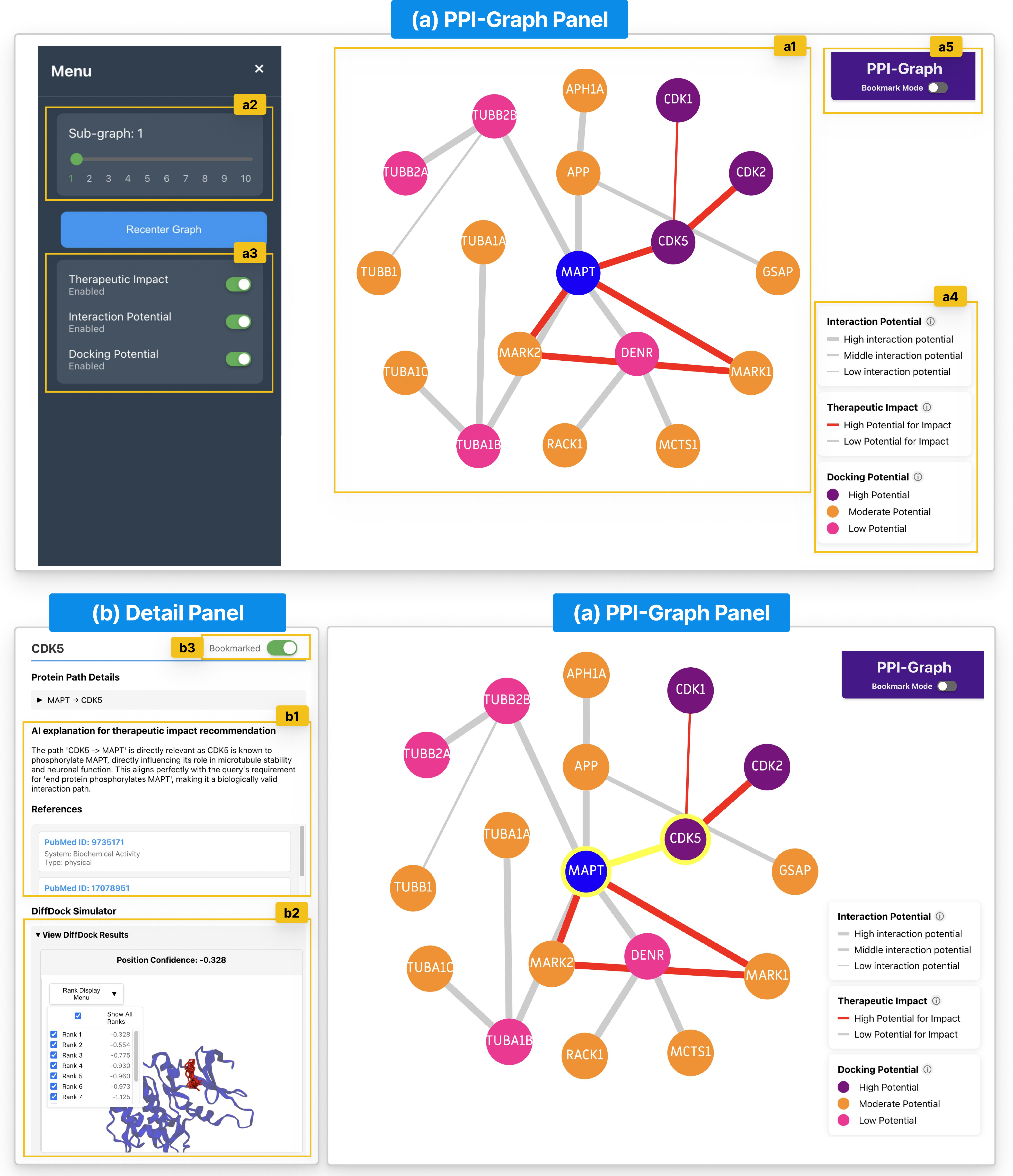}
    \caption{Overview of \system. (a) \panone enables users to explore and validate PPIs across three criteria: C1)  interaction potential, C2) therapeutic impact, and C3) docking potential. (b) \pantwo provides supporting information to enable users to justify AI suggestions, including references and 3D docking simulations.}
    \label{fig:HAPPIER_fin}
\end{figure*}

\section{\system} \label{HAPPI}
Building on design goals, we propose \system, an AI-powered tool that enables medicinal chemists to efficiently explore and verify hundreds of PPIs across all criteria (C1-C3) on integrated graph visualization. \system provides two main panels: 1) \panone: enables users to efficiently explore and verify proteins in a single integrated graph component showing multi-criteria satisfaction; and 2) \pantwo: allows users to validate AI suggestions with supporting information related to domain knowledge.

\subsection{\panone}\label{V1}
% \xac{let's avoid calling it 'visualization'; it's just a ui}
\panone offers an integrated search across the three criteria within a single graph-based component (Figure~\ref{fig:HAPPIER_fin}-a).
With \panone, users can 1) explore multiple subgraphs of hundreds of PPIs centered on their input protein without complexity from navigating a large-scale PPI graph, and 2) efficiently identify candidate PPIs that satisfy all criteria within the integrated graph without cognitive load from switching multiple interfaces. 
After exploring and validating PPIs, users can bookmark potential PPIs for later review. We employed AI models for therapeutic impact (C2) and docking potential (C3) (Please see the details of the models in {\S}\ref{V12}).

The use of \panone follows the steps outlined below:

\begin{itemize}
    \item Step 1: Users explore the first sub-graph centered on the initial protein that users input (Figure~\ref{fig:HAPPIER_fin}-a1).
    \item Step 2: Users select and switch between multiple subgraph views through the Sub-graph slider (Figure~\ref{fig:HAPPIER_fin}-a2).
    \item Step 3: Users apply criteria within the graph (Figure~\ref{fig:HAPPIER_fin}-a3). Users can not only apply all three criteria simultaneously but also selectively focus on a specific criterion of interest.
    \item Step 4: Users explore PPIs on the graph with the legend (Figure~\ref{fig:HAPPIER_fin}-a4), where functional and physical interactions (C1) are encoded through edge thickness, therapeutic impact (C2) is visualized by edge color, and docking potential (C3) is visualized by node color. You can check the details of the legend in {\S}\ref{lg} in the appendix.
\end{itemize}

Through these steps, we expect that users will have opportunities to iteratively expand and refine the scope of potential PPI candidates efficiently. 

\subsubsection{AI models for \panone}\label{V12}
\panone allows users to search for a therapeutic impact for C2 and docking simulation for C3 with AI models. 
For the therapeutic impact search, we employed a knowledge-graph-based Retrieval-Augmented Generation (RAG) framework~\cite{li2025grappi}.
The framework receives as input an initial protein (e.g., MAPT) and a desired therapeutic impact (e.g., to phosphorylate MAPT). It first extracts proteins connected to the initial protein and related information, such as protein description, papers, and interaction potential, from the database of STRING~\cite{szklarczyk2023string}. This framework constructs subgraphs ranked by their potential for interaction. 
These subgraphs are then decomposed into multiple paths, for which the framework generates explanations of therapeutic relevance with the retrieved information, corresponding protein descriptions, and papers. Finally, each pathway (PPI) is assigned a score ranging from 0 to 100, where higher scores indicate greater therapeutic potential.

For the docking simulation, we applied DiffDock~\cite{corso2022diffdock}, which is a state-of-the-art AI model specifically designed for predicting the binding modes of ligands to protein targets.
This model enables scientists to simulate where a ligand docks with a protein (docking site) and how the ligand docks with a protein (docking pose), which is the most important information when researchers understand the potential efficacy of a drug candidate docking with a protein. We applied the NVIDIA BioNeMo API~\cite{bionemo2024} that predicts binding affinity ($-15\text{--}0$), where higher scores indicate greater docking potential.

\begin{figure*}
    \centering
    \includegraphics[width=0.7
    \linewidth]{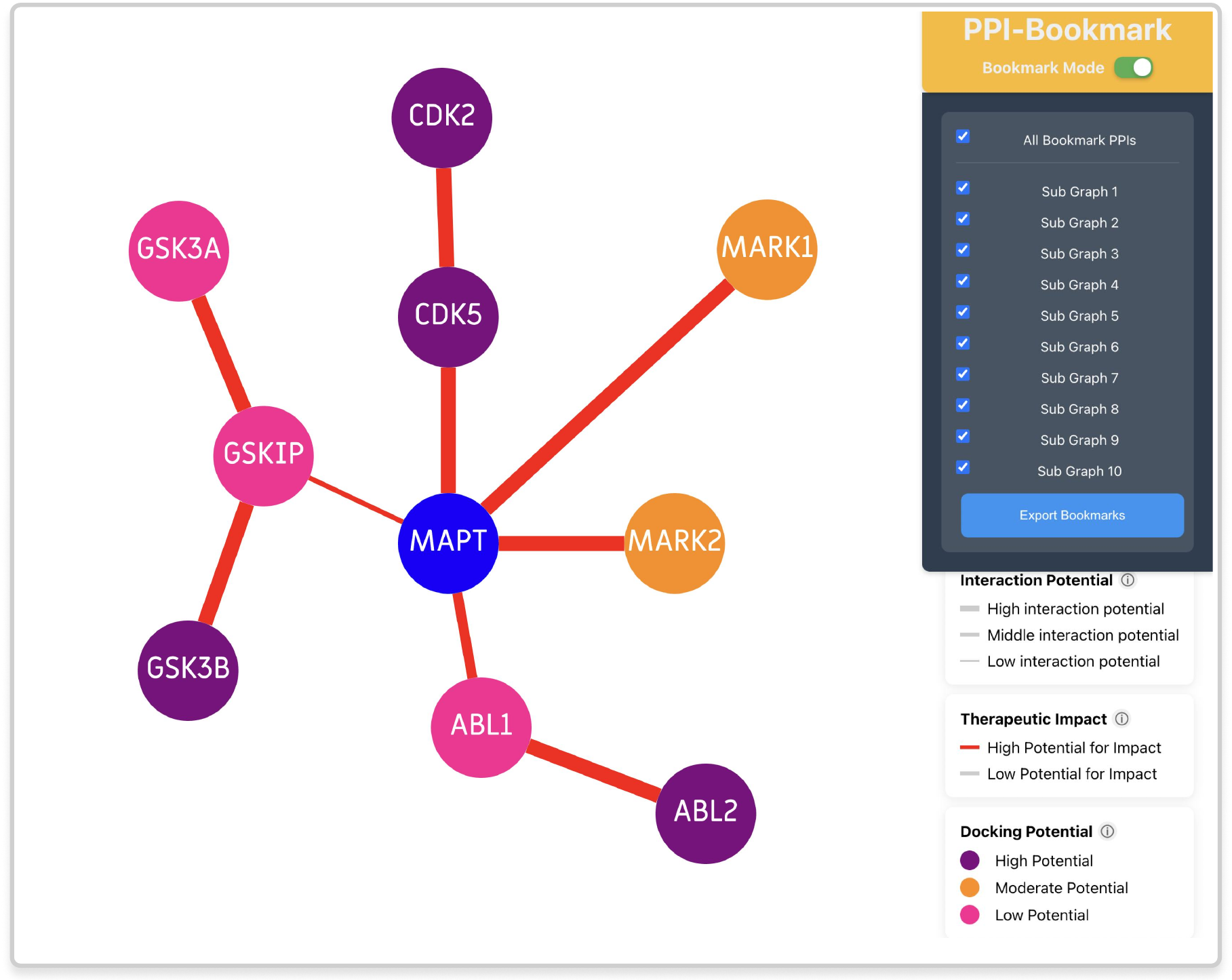}
    \caption{PPI-Bookmark mode enables users to build a personalized PPI graph that consists of bookmarked PPIs. This allows users to filter bookmarked PPIs by subgraph using the checkboxes on the right. }
    \label{fig:Subcomponent}
\end{figure*}

\subsection{\pantwo}\label{V2}
\pantwo enables users to justify AI suggestions with two types of domain knowledge information: (1) papers corresponding to AI-generated explanations and (2) multiple docking predictions from docking models. 

The use of \pantwo follows the steps outlined below:

\begin{itemize}
    \item Step 1: \pantwo opens when users click a specific PPI on the graph. 
    \item Step 2: Users check the explanations for the therapeutic impact recommendation for the PPI and papers to enable users to validate AI with domain knowledge (Figure~\ref{fig:HAPPIER_fin}-b1).
    \item Step 3: Users check the 3D docking simulation between the ligand reference that users input and the protein in the PPI. This enabled users to verify docking AI by checking for consistent binding of ligands near the protein folding site (Figure~\ref{fig:HAPPIER_fin}-b2).
    \item Step 4: Users bookmark the PPI if all supporting information makes sense by using a bookmark toggle (Figure~\ref{fig:HAPPIER_fin}-b3). This allows users to save selected PPIs directly into the PPI-bookmark (Figure~\ref{fig:Subcomponent}). 
\end{itemize}

Through these steps, we expect users to have opportunities to justify AI suggestions, facilitating the efficient refinement of the scope of potential PPI candidates. 

\subsubsection{\ftwo}\label{V2}
\ftwo enables users to store and compare selected PPIs that they bookmarked in \pantwo with information related to the three criteria (Figure~\ref{fig:Subcomponent}). The Bookmark Mode button (Figure~\ref{fig:HAPPIER_fin}-a5) enables users to view their own personalized PPIs consisting of bookmarked PPIs.
If users click particular PPIs, \system allows them to open \pantwo. \ftwo allows users to filter bookmarked PPIs by subgraph when users need to focus on specific PPIs. Through this process, users can finalize their curated PPIs.

%% file: text/06_User_Study.tex
\section{User study}\label{firstUS}
Our experiment had two research purposes. The first purpose was to validate whether \system supports divergent and convergent thinking in Target ID and achieves better results in hypothesis generation compared to existing alternative approaches (RQ1). The second purpose was to examine whether engaging in divergent and convergent thinking in \system leads to an increase in the perceived confidence in the output PPIs (RQ2).We conducted a between-subjects study in which participants were assigned to either the experimental or control group for a designed Target ID task.

\subsection{Participants}
We recruited ten professional medicinal chemists who have experience participating in AD-related projects and are currently working across ten different drug discovery research labs. These participants were newly recruited for the \system evaluation, distinct from the formative study. Table~\ref{tab:demo_p} in the appendix shows the participants' demographic information. Their career experience ranged from 4 to 23 years (mean=11.50, SD=7.06). To recruit participants from more diverse research institutes, we recruited people through online advertisements and snowball sampling. Our study was approved by the Institutional Review Board (IRB), and the consent of the participants was sought before the study. Each participant received a \$100 gift certificate. When reporting interview quotes, we use $P^{X}_{p}$ to denote participant number \textit{X} in this pilot study

\subsection{Procedure}\label{study}
We employed a between-subjects design. The experimental group was asked to use \system, while the control group, based on insights from the formative study, was instructed to use three existing interfaces: STRING for C1, Google Search for C2, and SwissTargetPrediction for C3.
Participants were asked to conduct a 30-minute task: to identify PPI candidates (hypothesis) suitable for wet-lab validation while satisfying three criteria, given the input information (initial protein: MAPT, therapeutic impact: to phosphorylate MAPT, and ligand: Roscovitine\footnote{https://bit.ly/3yXYyZJ}).
This task serves as a representative Alzheimer’s disease (AD) use case, as MAPT and Roscovitine are widely studied in the AD context. The professionals who participated in the formative study confirmed the appropriateness of this task design.

The study session proceeded as follows:
\begin{itemize}
    \item Step 1: The participants provided demographic (age, institute, year of experience). 
    \item Step 2: The participants were asked to complete the task in 30 minutes. 
    Participants in the experimental group conducted a 10-minute practice session after watching the instruction video on how to use \system
    \item Step 3: After the task, they were asked to answer the survey questions.
    \item Step 4: The participants were asked to be interviewed by the researchers about the degree of scientific discovery support offered by \system and its practical use.
\end{itemize} 

The study took place remotely using Zoom. Participants were asked to share their screen during the timed tasks and think-aloud, where participants were asked to verbally express their thoughts, decision-making processes, and difficulties while interacting with \system. This approach allowed us to observe participants' cognitive processes and problem-solving strategies. 

\begin{figure}
\centering
  \includegraphics[width=1\columnwidth]{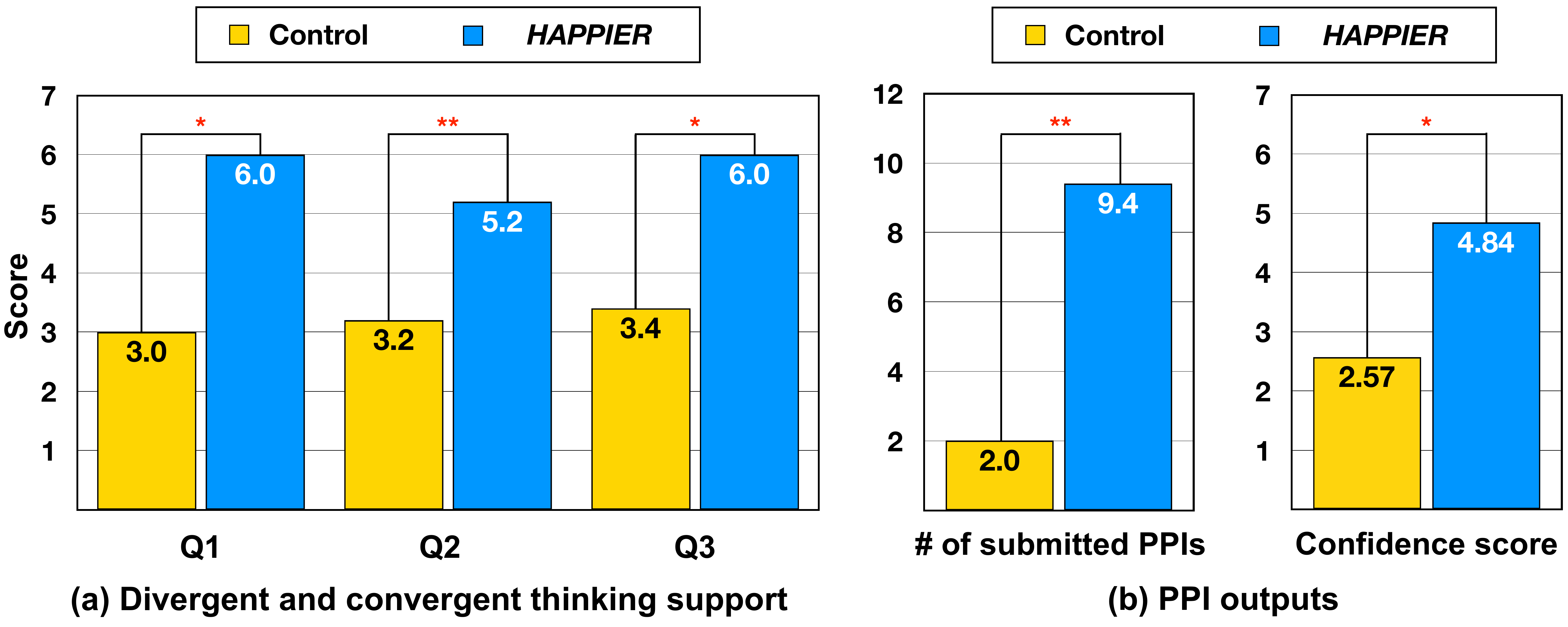}

  \caption{(a) The bar graph shows that the divergent and convergent thinking support was significantly higher in \system than in the control group. (b) The bar graph shows that the PPI outputs were significantly higher in the \system than in the control group ($^{*}p<0.05$, $^{**}p<0.01$).  }~\label{fig:results_Q1_Q2}
\end{figure}

\subsection{Measurements}\label{survey}

The survey questions were divided into two themes aligned with the purposes of this user study: 1) supporting divergent and convergent thinking and 2) achieving better PPI results than existing approaches.

The first set of questions focused on support for divergent and convergent thinking. Participants in the experimental group evaluate \system, while those in the control group assess their overall experience with the three interfaces. Users rated each task on the survey on a 7-point Likert scale (1 is ``Not at all'' and 7 is ``Very much''). The questions are as follows. 

\begin{itemize}
    \item Q1: To what extent does this interface extend the boundary of understanding PPIs by facilitating the exploration of multiple PPIs? (divergent thinking)
    \item Q2: To what extent does this interface constrain the boundary of understanding PPIs by assisting in evaluating PPIs? (convergent thinking)
    \item Q3: To what extent does this interface effectively support iterative cycles of exploration and verification of PPIs across the three criteria? (iterative cycles of divergent and convergent thinking)
\end{itemize}

The second set of questions evaluates the number and level of confidence in the PPIs that users submitted as the final output. Since the ultimate goal of \system is to help scientists create a list of PPI candidates for wet-lab experiments, both the number and confidence of PPIs are crucial for the next steps~\footnote{Validating PPIs through wet-lab experiments is often infeasible because the process is highly resource-intensive—each protein costs approximately \$1,400—and time-consuming, typically requiring several weeks to months to complete.}. 
Participants were asked to evaluate the confidence level of each submitted PPI on a 7-point Likert scale (1 = ``Not at all,'' 7 = ``Very much'').

%% file: text/08_Results.tex
\section{Results}\label{secondUS}

During the user study, we collected questionnaire responses from surveys, transcriptions from interviews, and 57 PPIs (experimental: 47; control: 10) submitted as final outputs. For RQ1 ({\S}\ref{Q1}), we employed a Wilcoxon rank-sum test~\cite{wilcoxon1945individual} to assess group differences in survey responses between experimental and control participants. For RQ2 ({\S}\ref{Q3}), we used a linear mixed effects model (LMER)~\cite{bates2015lme4} to examine differences in perceived confidence levels across three PPI categories based on their engagement in divergent and convergent thinking: (1) Both-DC (visited in both), (2) Either-DC (in one), and (3) Neither-DC (in neither). When reporting interview quotes, we use $P^{X}_{u}$  to denote participant number \textit{X} in the user study.

\subsection{Results for RQ1: Divergent–Convergent Thinking and PPI Outputs}\label{Q1}

\subsubsection{Iterative Cycles of Divergent and Convergent Thinking}
The three design goals of \system are to support 1) divergent thinking, 2) convergent thinking, and 3) iterative cycles of both divergent and convergent thinking. The results of the first set of questions indicate that the three design goals in \system are generally well supported. 
Figure~\ref{fig:results_Q1_Q2}-(a) shows that the scores of the experimental group (Q1: M=6.0, SD=0.71; Q2: M=5.2, SD=0.45; Q3: M=6.0, SD=0.71) are significantly higher than those of the control group (Q1: M=3.0, SD=1.87; Q2: M=3.2, SD=1.30; Q3: M=3.4, SD=1.52) across all three measures: Q1 ($U=24.5$, $p=0.014$), Q2 ($U=25.0$, $p=0.009$), and Q3 ($U=24.5$, $p=0.014$).
These results mean \system can support iterative cycles of divergent and convergent thinking in Target ID, highlighting the importance of integrating all criteria in a single tool in supporting the cycles.

\subsubsection{PPI outputs}
\system enabled scientists to generate more PPI outputs while maintaining their confidence level. 
Figure~\ref{fig:results_Q1_Q2}-(b) shows significant differences between the experimental and control groups. The number of submitted PPIs was higher in the experimental group (M=9.4, SD=3.51) than in the control group (M=2.0, SD=0.71) ($U=25.0$, $p=0.009$). The confidence level was also higher in the experimental group (M=4.62, SD=0.56) than in the control group (M=2.57, SD=1.51) ($U=24.0$, $p=0.021$).
Participants generated a larger number of higher-confidence hypotheses compared to the baseline, indicating that \system enhances the potential for generating more hypothesis candidates with strong confidence.

\begin{figure*}
\centering
  \includegraphics[width=2\columnwidth]{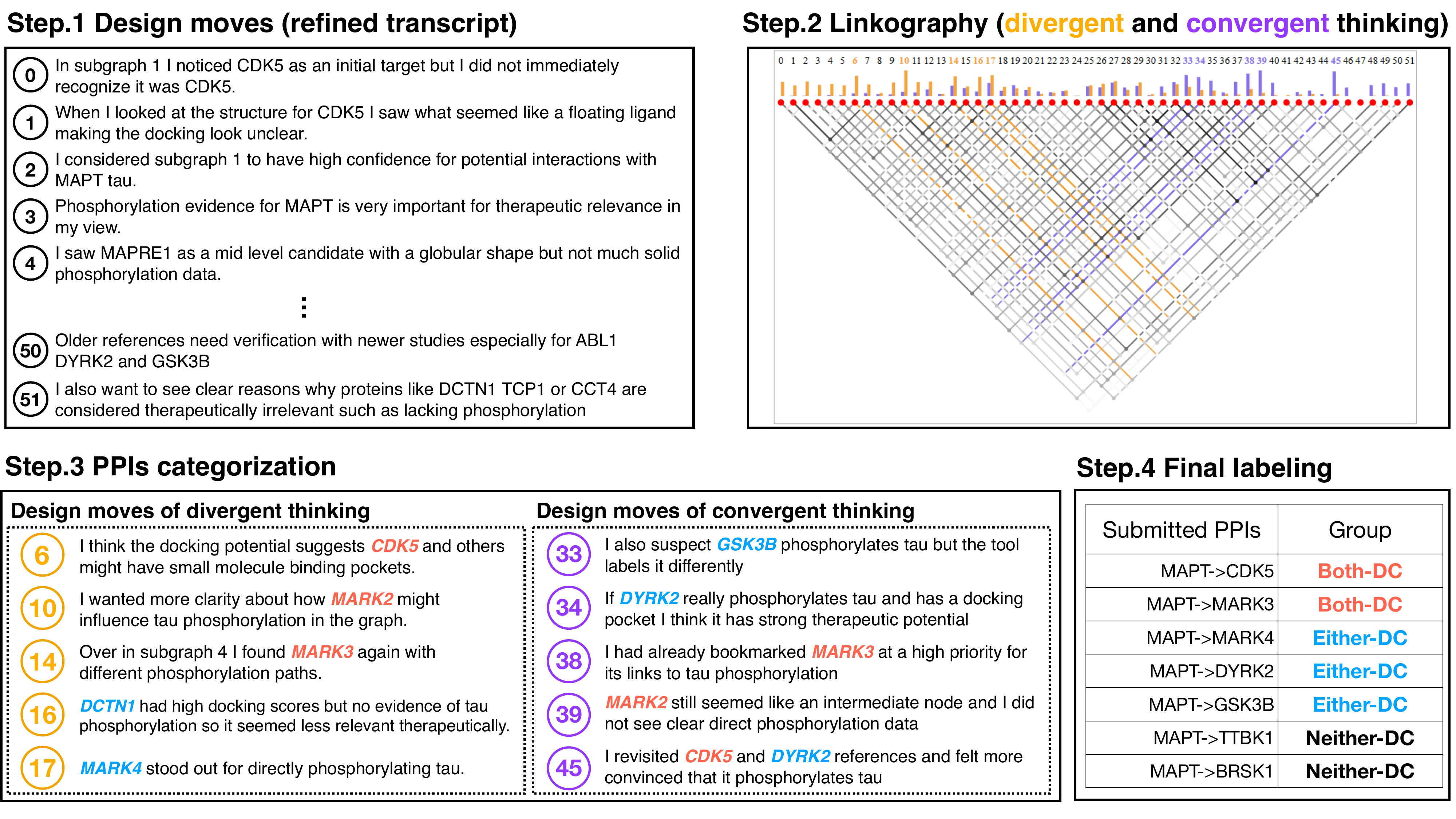}
  \caption{Process of categorizing the submitted PPIs. Step-1: generating design moves (DMs) by refining transcript in user studies; Step-2: making a linkography with the DMs to identify the DMs of \underline{D}ivergent (orange) and \underline{C}onvergent (purple) thinking; Step-3: identifying PPIs in the DMs of divergent and convergent thinking; Step-4: labeling the submitted PPIs into three groups: (1) Both-DC: present in both divergent (D) and convergent (C) thinking, (2) Either-DC: present in either one but not both, and (3) Neither-DC: present in neither.
  We considered PPIs that appeared in the DMs but were not among the submitted PPIs (e.g., MARK2 and DCTN1) as those that experts had filtered out through a process of divergent and convergent thinking.}
  ~\label{fig:linko}
\end{figure*}

\subsection{Results for RQ2: Engagement in Divergent–Convergent Thinking and Confidence}\label{Q3}
To determine whether engaging in the iterative cycle of divergent and convergent thinking leads to an increase in the perceived confidence in the output PPIs (RQ2), the submitted PPIs were categorized into three groups according to their association with iterative cycles, and we analyzed the relevance and confidence level within each group.
There are three steps:  
First, we identified divergent and convergent thinking processes while using \system using linkography ({\S}\ref{idct}), which is a method that analyzes connections between ideas and
enables the observation of divergent and convergent thinking.
Second, we categorized three groups based on their presence in divergent and convergent thinking: 
(1) Both-DC: present in both divergent (D) and convergent (C) thinking, (2) Either-DC: present in either one but not both, and (3) Neither-DC: present in neither ({\S}\ref{idct}).  
Third, we examined the statistical significance of relevance and confidence scores across the three groups ({\S}\ref{RCS}). Overall, we found that Both-DC yielded significantly higher relevance scores than Neither-DC, and higher confidence scores than Either-DC and Neither-DC.

\subsubsection{Identifying PPIs visited during divergent and convergent thinking phases}\label{idct}
As a first step, we identified divergent and convergent thinking in \system, using fuzzy linkography~\cite{smith2025fuzzy}, which is an AI framework to construct a linkography~\cite{goldschmidt2014linkography} automatically. There are four steps (Figure~\ref{fig:linko}: 1) generating design moves (DMs), 2) making a linkography, 3) categorizing PPIs, and 4) final labeling. 

We generated \textit{design moves} in chronological order, typically at the sentence level (Figure~\ref{fig:linko}-Step.1).
% \xac{how many design moves in total across all participants?}  
Following previous work~\cite{goldschmidt2014linkography, smith2025fuzzy}, we used the participants' transcripts, which were recorded while using \system. We created a final transcript by comparing two outputs of two audio transcription tools, Zoom~\footnote{https://www.zoom.com/} and Otter.ai~\footnote{https://get.otter.ai}. During this process, we selectively included only the content related to the design progress~\cite{goldschmidt2014linkography}. A total of 218 design moves were generated by five participants (mean: 43.6, SD: 4.4).

To make a linkography, we used fuzzy linkography, establishing a link automatically(Figure~\ref{fig:linko}-Step.2). In the system, links are generated when the cosine similarity — a continuous value between 0 and 1 — exceeds a predefined threshold (hyperparameter).
Our threshold was 0.75, following the default of fuzzy linkography, which proved its promising performance in tasks~\cite{smith2025fuzzy}. Fuzzy linkography identifies DMs by analyzing the top-\textit{k} frequent links: divergent thinking was defined as the top-\textit{k} most frequent forward links that connect a move to later moves it influences, while convergent thinking was defined as the top-\textit{k} most frequent backward links that connect a move to earlier moves from which it draws influence. For \textit{k}, we used approximately 10\% of the total design moves~\cite{goldschmidt2014linkography}. If the total number of moves is 52, then \textit{k} is set to 5 by rounding. Figure~\ref{fig:five_linkos} in the appendix shows five linkography diagrams of five participants ($P^{1}_{u}$-$P^{5}_{u}$). Through these steps, we identified the DMs of divergent (\textit{D}) and convergent (\textit{C}) thinking.  
 
We categorized all PPIs into three groups (e.g., Both-DC, Either-DC, and Neither-DC) based on their presence in divergent (\textit{D}) and convergent (\textit{C}) thinking (Figure~\ref{fig:linko}-Step.3). Five participants submitted a total of 47 PPIs (mean = 9.4, SD = 2.3). We finally labeled 47 submitted PPIs into three groups: 9 in the Both-DC, 10 in the Either-DC group, and 28 in the Neither-DC (Figure~\ref{fig:linko}-Step.4).

% \begin{itemize}[leftmargin=1.5em]
%   \item \textbf{Both}: PPIs involved in both types of thinking:
%   \[
%   \text{Both-DC} = \{ p \in P \mid p \in C \land p \in D \}
%   \]
%   where \( p \) denotes a single protein-protein interaction (PPI), and \( P \) is the set of all submitted PPIs.

%   \item \textbf{Either}: PPIs involved in exactly one type of thinking (either convergent or divergent, but not both):
%   \[
%   \text{Either-DC} = \{ p \in P \mid p \in C \triangle D \}
%   \]
%   where \( \triangle \) denotes the symmetric difference between sets.

%   \item \textbf{Neither}: PPIs not involved in either type of thinking:
%   \[
%   \text{Neither-DC} = \{ p \in P \mid p \notin C \land p \notin D \}
%   \]
% \end{itemize}

% \xac{i think readers can understand the 3 categories without 9.3.2.}

\begin{figure}
\centering
  \includegraphics[width=0.77\columnwidth]{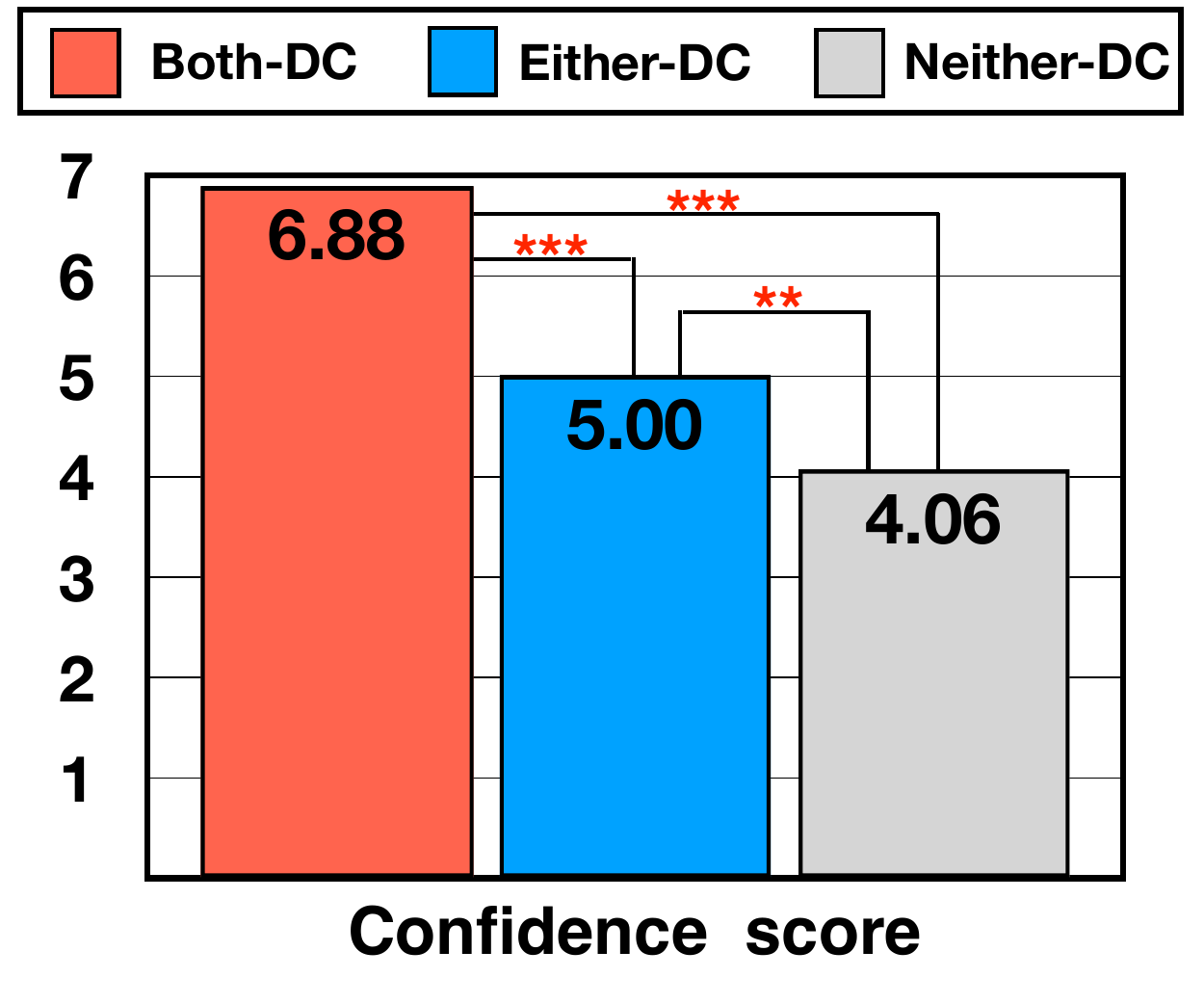}

  \caption{The bar plot showing the significant differences in confidence score among three groups: (1) Both-DC: present in both divergent (D) and convergent (C) thinking, (2) Either-DC: present in either one but not both, and (3) Neither-DC: present in neither ($^{**}p<0.01$, $^{***}p<0.001$). 
  % \xac{need to define the three groups. each fig needs to be self-contained without viewing other figs}
  % \xac{label relevance and confidence on the x axes}
  }
  ~\label{fig:RC_scores}
\end{figure}

\subsubsection{Confidence score}\label{RCS}
Given the repeated measures design and the need to account for participant-level variability, we employed a linear mixed effects model (LMER)~\cite{bates2015fitting} to determine the statistical significance of the confidence score between the three groups (e.g., Both-DC, Either-DC, Neither-DC). The dataset included participant identifiers (e.g., $P^{1}_{u}$-$P^{10}_{u}$), submitted PPIs (e.g., CDK5–MAPT, MARK2–MAPT), group labels (e.g., Both-DC, Either-DC, Neither-DC), and the corresponding confidence score for each PPI.  
% \xac{at some point we need to report how many PPIs were submitted per participant and in total.} 

We set the confidence score as the dependent variable, respectively. The group (Both-DC, Either-DC, Neither-DC) and the participants were included as a fixed effect and a random intercept, respectively.Figure~\ref{fig:RC_scores} indicates confidence in the submitted PPIs was significantly higher for PPIs associated with both convergent and divergent thinking. The \textit{Both-DC} group showed significantly higher confidence scores than the \textit{Either-DC} group ($\beta = -1.886$, $p < .001$) and the \textit{Neither-DC} group ($\beta = -2.829$, $p < .001$). The minimal variance of the random effect for participant ($\sigma^2 = 0.008$) suggests that individual differences had little impact. A post-hoc Tukey HSD test confirmed that all pairwise comparisons were statistically significant: \textit{Both-DC} vs. \textit{Either-DC} ($p < .001$), \textit{Both-DC} vs. \textit{Neither-DC} ($p < .001$), and \textit{Either-DC} vs. \textit{Neither-DC} ($p = .0116$). These results suggest that engaging experts in both divergent and convergent thinking can significantly enhance the perceived confidence in PPIs (RQ2), highlighting the importance of engaging in both divergent and convergent thinking in supporting hypothesis generation.

% \xac{end with a brief qualitative statement that connects the above findings to RQ2. implications for design will be discussed in the next sec?}

\subsection{Interview Findings}\label{interview}
During the interview, the participants explained in detail how \system supports divergent and convergent thinking in Target ID. After completing the interviews, we applied thematic analysis and iterative open coding~\cite{clarke2015thematic} to analyze the interview transcripts.

\subsubsection{Integration of Criteria Supports Divergent–Convergent Thinking (RQ1)}\label{itv_DC}
All participants commented that \system provided sufficient support in promoting iterative cycles of divergent and convergent thinking in Target ID. They could efficiently explore and validate PPIs across three criteria on a graph-based component, which leads to the iterative process of divergent and convergent thinking.  In particular, \system enabled them to uncover previously inaccessible PPIs in subgraphs and strategically narrow down their shortlist of candidates by integrating therapeutic impact search and simultaneous docking simulations within the subgraph. For example, \textit{``I didn't find the TUBB in STRING, but it appeared in \foneone. I could quickly validate TUBB with \fonetwo and \fonethree, and I chose it''} ($P^{2}_{u}$). 
\textit{``I usually spend a lot of time on Google for understanding a protein. General understanding of multiple proteins at once has been incredibly helpful''} ($P^{3}_{u}$). 
\textit{``Docking multiple proteins on a graph simultaneously is extremely efficient. When using SwissTargetPrediction, I should validate PPIs one by one''} ($P^{1}_{u}$).
These accounts highlight the importance of integrating multiple criteria in supporting the cycles of divergent and convergent thinking, which is essential in supporting hypothesis generation. 

\subsubsection{Iterative Cycles Drive Better Outputs (RQ2)}
All participants mentioned that the iterative cycles of divergent and convergent thinking in \system help increase confidence level for their PPI outputs. They were able to make more scientifically grounded judgments from their initial intuition. Although they initially felt confident about a particular PPI, comparing it with other PPIs that have either better or worse conditions led them to adjust their initial intuition. This process also helped increase their confidence in the final outcome. For example, \textit{``At first, I could tell that CDK5 met all three criteria. And the more I compared it with other proteins, the more it became clear that CDK5 was the strongest candidate. I think that’s probably why I said I was most confident about CDK5''} ($P^{4}_{u}$). 
\textit{``I initially thought MARK4 looked really promising. But after seeing other PPIs with much stronger properties, I started to feel a bit less confident about MARK4''} ($P^{5}_{u}$). 
These accounts demonstrate the positive relationship between engaging in the iterative cycles of divergent and convergent thinking and perceived confidence in hypothesis output, accounting for the underlying mechanism of \system.

\subsubsection{Intuitive Design of \system: from Overview to Details}
Participants said that using \system felt highly intuitive and easy to use. They mentioned that \system first allowed them to get an overview of the PPIs based on whether each satisfied the three key criteria, and then, when more detailed inspection was needed, they could access information for each PPI—such as explanation information with relevant papers and multiple docking simulation results. Given the complexity of the heterogeneous data types linked to the three criteria, participants noted that the absence of a user-friendly design could have easily resulted in cognitive overload. For example, \textit{``If all that information had come up at once, I think it would’ve been really hard to use. But the system made it easy by giving me a big-picture view first, and then letting me check the details only when I needed to''} ($P^{4}_{u}$). 
\textit{``When I normally do target identification, I sometimes have to open dozens of windows to verify the three criteria. It often gets overwhelming trying to process all that information. In contrast, \system makes it really easy to get an overview and intuitively verify relevant information. I think it would be incredibly helpful in the real world''} ($P^{3}_{u}$). 
These accounts suggest that the intuitive design of \system, which provides an overview before details, helps users generate hypotheses efficiently while mitigating excessive cognitive load from exposure to heterogeneous data.

\subsubsection{Challenges: Necessity for Domain Knowledge–Centered Interaction}
Participants expressed a need for more substantial support in two areas: (1) traceability for AI output: identifying which parts of the paper were relevant to the AI-generated suggestions, and (2) Human-in-the-loop input: viewing a more robust version of the system that allows them to incorporate their knowledge or upload custom data. For the traceability, participants wished to go beyond just seeing related papers—they wanted to know exactly which parts supported the AI-generated outputs. For example, 
\textit{``If I could see exactly which part of a paper was related to the AI suggestion, I think I could verify things much more efficiently. It takes a lot of time to read through the entire paper''} ($P^{5}_{u}$). For the human-in-the-loop input, participants wished to upload their data and knowledge to have a better version of the system. \textit{``As far as I know, the APP protein is known to be strongly related to the phosphorylation of MAPT. But the AI suggestion didn’t really emphasize that connection. I feel like if I could update the system with my own papers and knowledge, it could generate much stronger results''} ($P^{1}_{u}$). Together, these underscore the importance of supporting domain knowledge–centered interaction in both input and output in human-AI collaboration for hypothesis generation.

%% file: text/09_Discussion.tex
\section{Discussion}
We summarize the findings of the study and discuss their implications, and then we discuss the present work’s limitations and our plans for future research.

\subsection{Lessons Learned for Future AI-powered Hypothesis Generation Supporting Tools Design}

\subsubsection{Supporting Iterative cycles of Divergent and Convergent thinking for Hypothesis Generation}
In this research, we identified the importance of supporting iterative cycles of divergent and convergent thinking in facilitating hypothesis generation in drug discovery. Exploring PPIs across multiple subgraphs enables users to expand the hypothesis space by moving from well-studied to less-studied but potentially novel PPIs, while verifying PPIs with AI models for therapeutic impact and docking simulations allows them to narrow the space by validating hundreds of candidates simultaneously.
By integrating three criteria for Target ID into a single graph, \system streamlines iterative cycles of exploring and verifying hypotheses, allowing users to adjust standards for generating reasonable hypotheses.

This approach may have the potential to be applied in designing AI-powered tools for other health domains, such as bioinformatics, clinical decision support, and public health, which require hypothesis generation with multiple criteria. \system can be applied to other domains by following the three steps. First, future work needs to consider the interactive medium to expand the hypothesis scope (e.g., from a larger graph to multiple subgraphs). Second, within this medium, AI models should help users efficiently validate hypotheses (e.g., therapeutic impact search and docking simulation) across criteria with supporting information (e.g., reference and multiple docking simulations). Third, within this medium, future work should enable them to explore and verify hypotheses iteratively (e.g., showing multi-criteria satisfaction).

\subsubsection{Roles of Engaging in Iterative Cycles}
Our study demonstrates the relationship between engaging in divergent and convergent thinking and scientists’ perceived confidence in the output.
Previous research suggests that confidence in new ideas increases through cycles of divergent and convergent thinking, as divergence reassures individuals that possibilities have been broadly explored, while convergence justifies that selected ideas are evidence-based~\cite{cropley2006praise,finke1996creative}.
These insights illustrate the hypothesis generation process supported by our integrated interface, which enables medicinal chemists to iteratively explore and refine hypotheses across multiple criteria, resulting in a larger number of hypotheses generated with higher confidence.

Furthermore, making it transparent how each output was cognitively engaged—in Both, Either, or Neither of the divergent and/or convergent processes may help users recognize missed opportunities, validate their reasoning, and consider alternative research paths. For example, at the end of a session, AI can provide an overview of the scientific discoveries with indicators of divergent and convergent thinking engagement. With this information, researchers can revisit outcomes they consider plausible but remain uncertain about, enabling more confident verification and refinement in hypothesis generation.
Furthermore, AI could proactively recommend detailed information on outputs in the Either and Neither groups that require further exploration. This enables promoting mutual understanding between humans and AI beyond one-way support, aligning more closely with the essence of human-AI collaboration~\cite{fugener2022cognitive} in other health domains. 
For instance, in public health~\cite{jungwirth2023artificial, olawade2023using}, when medical researchers first formulate hypotheses, such as whether certain lifestyle factors (e.g., diet, exercise) reduce specific diseases, a future AI-powered system can proactively suggest revisiting hypotheses that show potential but lack sufficient engagement in the iterative cycles based on their engagement data.

\subsubsection{Rich Information vs. Information Overload} 
Our finding highlights that \system enables users to explore and verify complex heterogeneous data intuitively and efficiently.
In the context of drug discovery, providing a greater diversity and quantity of information can enhance the potential for novel hypothesis generation, but it may also increase the risk of information overload~\cite{zuech2015intrusion}. When users are presented with excessive heterogeneous information at the same time, they can become overwhelmed, resulting in increased cognitive load~\cite{harper2009toward}.
This implies that exploring and verifying hypotheses across heterogeneous data within a single platform is vital for hypothesis generation, but also raises the risk of cognitive overload.

\system was designed to help users grasp an overview of the PPI candidates first, and then explore details on demand, enabling users to explore and verify complex, heterogeneous data seamlessly.
Previous works’ findings align with this point. Shneiderman proposed the Visual Information Seeking Mantra~\cite{shneiderman2003eyes}, which advocates for presenting users with a high-level overview first, followed by zooming and filtering, and finally providing details on demand. This approach enables users to efficiently explore, analyze, and interpret complex datasets without being overwhelmed. Similarly, Nielsen introduced Progressive Disclosure, a technique that initially presents only essential information and gradually reveals more details as needed, allowing users to absorb information in manageable steps~\cite{nielsen2006progressive}. Our study emphasizes the utility of heterogeneous data for hypothesis generation, while also addressing the potential risk of information overload and proposing possible solutions for future work.

\subsubsection{Domain Knowledge-based Interaction for Human-AI Collaboration}

Our findings demonstrate the importance of supporting domain knowledge–based interaction for human–AI collaboration. On the output side, participants sought to ensure that AI-generated results were consistent with domain knowledge. For instance, in the therapeutic impact search (C2), they justify the AI’s explanations by examining which sections of the related papers were considered. On the input side, participants wanted to reflect their professional knowledge into the system to deliver their idea on AI. For example, they expressed a desire to incorporate their own expertise or upload custom data, noting that such human-in-the-loop input could correct gaps in AI reasoning.
Previous works’ findings align with this point. Xie et al~\cite{xie2020chexplain} proposed CheXplanin, a system that enables physicians to understand AI-enabled chest X-ray analysis by interaction based on domain knowledge. Cai et al.~\cite{cai2019human} highlighted that providing experts with opportunities to refine AI predictions—such as emphasizing critical parts of the model’s output based on their own criteria—can influence trust in decision-making.
In this way, supporting domain knowledge–grounded interaction represents a key design implication for future research on generating promising hypotheses.

\subsection{Limitations and Future Work}
First, evaluating the quality of PPI output from \system based on the number and confidence of submitted PPIs is not entirely sufficient. The only way to fully assess the quality of \system's output is through scientific validation of the predicted PPIs via wet-lab experiments. However, due to the high cost—approximately \$1,400 per protein—there are practical limitations to performing such experiments. In future work, we aim to acquire experimentally validated datasets to serve as ground truth and quantitatively evaluate target proteins identified via \system.

Second, the time of the user study (30 minutes) and the number of participants might limit the generalizability of our findings. Participants discovered multiple PPIs with high confidence within the limited time. However, since identifying PPI candidates in Target ID is a task that takes at least a month, caution is required when interpreting the results. In addition, although a total of 15 participants were involved throughout the three studies, the sample may not fully represent the drug discovery domain. We believe broadening collaboration with more drug discovery labs could be the solution. In future research, we aim to enhance the practical applicability of \system through a user study that spans at least a week with more professionals to enhance the generalizability of our findings.

%% file: text/10_Conclusion.tex
\section{Conclusion}
Our study proposed a new methodology to support hypothesis generation in Target ID for human health. Through integrating all criteria for Target ID into a single platform, \system supports iterative cycles of divergent and convergent thinking in hypothesis generation. Both the quantitative and qualitative results of our user studies confirmed that \system can facilitate the iterative cycles of divergent and convergent thinking and enables generating more high-confidence hypotheses; further, there is a positive relationship between engaging in such cycles and confidence in hypothesis output. Our findings suggest design implications for AI-powered tools in scientific discovery, particularly in promoting proactive hypothesis generation grounded in engagement in iterative cycles and in contextualizing human–AI interaction through domain-specific knowledge.

%% file: text/Appendix.tex
\section{Appendix}
\input{table/Rel_interface.tex}

\begin{center}

\input{table/participants-main}

\end{center}

\begin{center}
    \input{table/formative}
\end{center}

\subsection{Legend for \system}\label{lg}
In both the PPI-graph mode and the PPI-Bookmark mode, the legend is provided for these graph modes. This provides the 3 different legends labeled with meanings for line thickness, line color, and node color. The three legends are referred to as the Interaction Potential, the Therapeutic Impact,  and the Docking Potential, respectively.

The Interaction Potential shows the possibility of the two proteins interacting with each other and is represented by a thickness of the edges, with the thickest being the highest potential and the thinnest being the lowest potential. These potentials are based on the combined score from STRING. The combined score is a combination of the physical attributes of the proteins and the interaction between the two proteins. This combined score outputs in the range of 0 to 1000. The thickness of the lines is based on each third of the combined score outputs. The 0 to 333 uses the thinnest line, a value of 333 to 666 uses the medium-thickness line, and a value of 667 to 1000 uses the thickest line. 

% \jay{@Chris, can you explain with the revised version that use the combined score?}\chris{added it in}
% value of two proteins relationship score provided by GraPPI. The value ranges from 0 to 999. A value of 0 to 333 uses the thinnest line, a value of 333 to 666 uses the medium thickness line, and a value of 666 to 999 uses the thickest line.

The Therapeutic Impact shows the possibility that the PPI pathway has a therapeutic impact and is represented by the color of the edges, with red being high potential and gray being low potential. Line coloring is based on the pathway score provided by the RAG framework in \label{LLM-1}. The edge color is active if the pathway score is above 0. 
% \jay{is there explanation for the pathway score? if so, use the same word?}\aaron{Adding the explanation of pathway scores in A.1. The words are the same.}
% \jay{if we used the name of GraPPI, our submission should be desk-rejected, because anonymity is not guaranteed}\aaron{Thanks for reminding. I have double checked it. There is no 'GraPPI' in the draft.}

The Docking Potential shows the potential of the ligand to dock onto the selected protein and is represented by the color of the node, with purple being the high potential, orange being the moderate potential, and pink being the low potential. These potentials are based on binding affinity (-15\~0) given from the Diffdock model~\cite{corso2022diffdock} from NVIDIA BioNeMo API~\cite{bionemo2024}. Purple is used for any node with a value above -0.5, orange is used for any node with a value between -0.5 and -2, and pink is used for any value below -2.

\begin{figure*}
\centering
  \includegraphics[width=2\columnwidth]{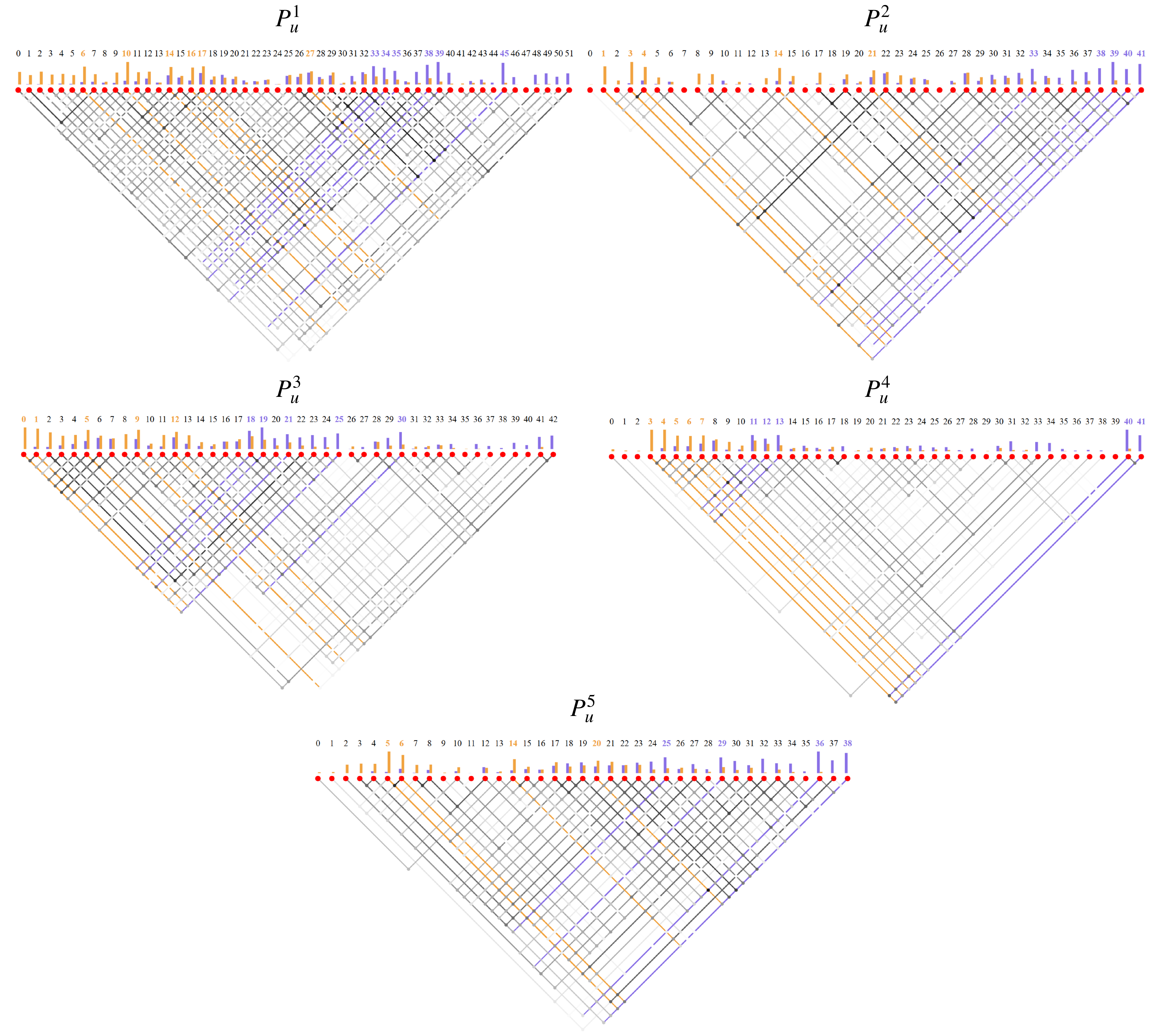}
    \caption{Linkography diagrams of five participants of the experimental group ($P^{1}_{u}$-$P^{5}_{u}$). Red nodes represent design moves, and the indices above them indicate their chronological order. Orange-colored indices and edges represent divergent thinking, while purple-colored indices and edges represent convergent thinking.  
    Links are established based on semantic or functional relevance, with edge color intensity indicating the degree of similarity—the darker the edge, the higher the similarity between connected moves. 
    }
  ~\label{fig:five_linkos}
\end{figure*}

%% file: table/Rel_interface.tex
\begin{table*}[h]
\centering
\caption{Web-based interface for target prediction (through interviews with four drug discovery professionals). These interfaces have only partially supported three criteria: C1) \textit{physical and functional interactions}, C2) \textit{therapeutic impact}, and C3) \textit{docking potential}. 
% \xac{move to related work?}
}
\label{tab:SDtool}

\begin{tabularx}{\textwidth}{p{2.5cm}|X|>{\centering\arraybackslash}p{2.5cm}|>{\centering\arraybackslash}c}
\toprule
\multicolumn{1}{c|}{Name} & \multicolumn{1}{c|}{Description} & \multicolumn{1}{c}{Criteria} & \multicolumn{1}{c}{Ref.}  \\

\midrule
\href{https://string-db.org/}{STRING}  & The STRING dataset is a comprehensive resource that integrates experimental data, text mining, and computational predictions to provide high-confidence protein-protein interaction networks  & C1  & \cite{szklarczyk2021string} \\

\midrule
\href{https://www.targetvalidation.org}{Open Targets Platform}  & The Open Targets Platform is an information-rich resource that offers evidence on the connections between known drug targets and diseases, facilitating the discovery and ranking of drug targets. & C1 & \cite{ochoa2021open} \\

\midrule
\href{https://scholar.google.com/}{Google Scholar}  & Google Scholar is an academic search engine that helps researchers find articles, including those on the therapeutic impact of proteins. & C2 & \cite{ochoa2021open} \\

\midrule
\href{https://www.disgenet.org}{DisGenNET}  & DisGeNET offers data on genes and genetic variants linked to human diseases.  & 
C2 &  \cite{pinero2020disgenet} \\

\midrule
\href{https://maayanlab.cloud/Harmonizome/}{Harmonizome}  & Harmonizome is a collection of comprehensive and processed knowledge gathered from over 70 major online resources on genes and proteins. & C3   &  \cite{rouillard2016harmonizome} \\

\midrule
\href{http://www.cbligand.org/TargetHunter}{TargetHunter}  & TargetHunter uses the TAMOSIC algorithm to efficiently predict the biological targets of compounds in question.  & C3   & \cite{wang2013targethunter} \\

\midrule
\href{https://sea.bkslab.org/}{Similarity Ensemble Approach (SEA)}  & SEA ranks target proteins by calculating a similarity score based on the chemical similarity of 65,000 ligands, which are grouped according to human protein targets.  & C3  & \cite{keiser2009predicting}   \\

\midrule
\href{http://www.swisstargetprediction.ch}{SwissTarget Prediction}  & 
SwissTarget Prediction uses a similarity search to predict potential drug targets for queried molecules, and the updated version includes 376,342 experimentally active compounds and 3,068 macromolecular targets.  & C3  & \cite{gfeller2014swisstargetprediction}  \\

\midrule
\href{https://prediction.charite.de}{SuperPred}  & 
SuperPred employs a linear regression model trained with ECFP4 fingerprints to predict the target proteins of compounds. & C3   & \cite{dunkel2008superpred}  \\

\midrule
\href{http://gdbtools.unibe.ch:8080/PPB}{Polypharmacology browser}  & 
The Polypharmacology Browser utilizes ten different fingerprints, molecular shapes, and substructure data to predict the most likely drug targets for a given small molecule. & C3   & \cite{awale2017polypharmacology}  \\

\midrule
\href{https://mips.helmholtz-muenchen.de/proj/hitpick}{HitPick}  & 
HitPick predicts potential drug targets for hit compounds using the B-score method, a one-nearest-neighbor similarity search, and a modified naïve Bayesian model.  & C3  & \cite{liu2013hitpick}  \\

\midrule
\href{http://moltarpred.marseille.inserm.fr}{MolTarPred}  & 
MolTarPred generates a list of potential drug targets and compounds that may have similar properties.  & C3  & \cite{peon2019moltarpred} \\

\midrule
\href{http://mussel.uniba.it:5000}{MuSSeL}  & 
MuSSeL employs a multi-fingerprint similarity search algorithm to predict the potential drug targets of small molecules.  &C3  & \cite{alberga2018new}  \\

\bottomrule
\end{tabularx}
\end{table*}

%% file: table/participants-main.tex
\begin{table*}[ht]
\centering
\small
\caption{Participant demographic information by group in the main user study. Ligand ID means ligand identification, where researchers identify ligands that bind to a target protein. }
\label{tab:demo_p}
\begin{tabular}{llcccccl}
\toprule
\textbf{Group} & \textbf{ID} & \textbf{Age} & \textbf{Gender} & \textbf{Profession} & \textbf{Domain} & \textbf{Work experience} \\
\midrule
\multirow{5}{*}{Experimental}
 & $P^{1}_{u}$ & 36--45 & Male & Senior Researcher & Target ID & 23 years \\
 & $P^{2}_{u}$ & 36--45 & Male & Senior Researcher & Target and Ligand ID & 18 years \\
 & $P^{3}_{u}$ & 26--35 & Male & Ph.D candidate & Target and Ligand ID & 4 years \\
 & $P^{4}_{u}$ & 26--35 & Female & Ph.D candidate & Target and Ligand ID & 4 years \\
 & $P^{5}_{u}$ & 36--45 & Male & Senior Researcher & Target ID & 14 years \\
\midrule
\multirow{5}{*}{Control}
 & $P^{6}_{u}$ & 26--35 & Male & Ph.D candidate & Target and Ligand ID & 5 years \\
 & $P^{7}_{u}$ & 26--35 & Male & Ph.D candidate & Target and Ligand ID & 4 years \\
 & $P^{8}_{u}$ & 36--45 & Female & Senior Researcher & Target and Ligand ID & 18 years \\
 & $P^{9}_{u}$ & 36--45 & Female & Postdoctoral Researcher & Target and Ligand ID & 10 years \\
 & $P^{10}_{u}$ & 26--35 & Male & Senior Researcher & Target ID & 15 years \\
\bottomrule
\end{tabular}

\end{table*}

%% file: table/formative.tex
\begin{table*}[ht]
\centering
\small
\caption{Participant demographic information by group in the formative study}
\label{tab:demo_f}
\begin{tabular}{lcccccl}
\toprule
\textbf{ID} & \textbf{Age} & \textbf{Gender} & \textbf{Profession} & \textbf{Domain} & \textbf{Work experience} \\
\midrule
$P^{1}_{f}$ & 26$\sim$35 & Male & Ph.D Candidate & Target ID & 7 years \\
$P^{2}_{f}$ & 26$\sim$35 & Male & Ph.D Candidate & Target ID & 8 years \\
$P^{3}_{f}$ & 36$\sim$45 & Female & Postdoctoral Researcher & Target ID & 9 years \\
$P^{4}_{f}$ & 36$\sim$45 & Male & Senior Researcher & Target ID & 18 years \\
$P^{5}_{f}$ & 26$\sim$35 & Male & Senior Researcher & Target ID & 15 years \\
\bottomrule
\end{tabular}

\end{table*}